\documentclass[12pt]{article}

\usepackage{lscape}
\usepackage{color,soul}
\usepackage{comment}
\usepackage{amsmath}
\usepackage{graphicx,psfrag,epsf}
\usepackage{enumerate}
\usepackage{xurl} 
\usepackage{amssymb, amsthm, amsfonts, bm, tabu, enumitem, xcolor, subcaption, booktabs, authblk}
\usepackage[linesnumbered,ruled,vlined]{algorithm2e}
\SetArgSty{textnormal}
\SetCommentSty{mycommfont} 
\SetKwInput{KwInput}{Input}                
\SetKwInput{KwOutput}{Output}              
\SetKwInput{KwNote}{Note}              
\usepackage{footnote}
\usepackage{animate}
\usepackage{caption}
\usepackage{xr-hyper}
\usepackage[colorlinks=true,linkcolor=blue,citecolor=blue]{hyperref}
\usepackage{multirow}
\usepackage{setspace}
\def\spacingset#1{\renewcommand{\baselinestretch}%
{#1}\small\normalsize} \spacingset{1}
\usepackage{natbib}
\usepackage[symbol]{footmisc}

\newcommand{\matern}{Mat\'{e}rn }
\newcommand{\R}{\mathbb{R}}

\newcommand{\vt}{\mathbf}
\newcommand{\mc}{\mathcal}

\graphicspath{ {./figure/} }

\newcommand{\articletitle}{Spatial predictions on\\physically constrained domains:\\Applications to Arctic sea salinity data}

\newcommand{\repourl}{\if0\blind
	{at \url{https://github.com/jinbora0720/boraGP}}\fi
	\if1\blind
	{on GitHub}\fi}

\newcommand{\repourlanal}{\if0\blind
	{at \url{https://github.com/jinbora0720/boraGP-sss}}\fi
	\if1\blind
	{on GitHub}\fi}
\newcommand{\blind}{0}
\newcommand{\standaloneappendix}{0}

\usepackage[hmargin=1.0in,vmargin=1.0in]{geometry}

\graphicspath{ {./figure/} }
\DeclareGraphicsExtensions{.pdf,.png} 

\if1\standaloneappendix
{
\makeatletter
\newcommand*{\addFileDependency}[1]{
  \typeout{(#1)}
  \@addtofilelist{#1}
  \IfFileExists{#1}{}{\typeout{No file #1.}}
}
\makeatother
\newcommand*{\myexternaldocument}[1]{%
    \externaldocument{#1}%
    \addFileDependency{#1.tex}%
    \addFileDependency{#1.aux}%
}
\myexternaldocument{appendix}
} \fi

\begin{document}

\title{\bf \articletitle}
    \if0\blind
	{ 
		\author[1,*]{Bora Jin}
		\author[2,3,4]{Amy H. Herring}
		\author[2,5]{David Dunson}
            \affil[1]{Department of Biostatistics, Johns Hopkins University}
		\affil[2]{Department of Statistical Science, Duke University}
		\affil[3]{Duke Global Health Institute}
		\affil[4]{Department of Biostatistics \& Bioinformatics, Duke University}
		\affil[5]{Department of Mathematics, Duke University}
		\affil[*]{email: bjin9@jh.edu} 
		\maketitle
	} \fi
	
	\if1\blind
	{
	\begin{center}
	{\spacingset{1} \LARGE\bf \articletitle}
	\end{center}
	\medskip
	} \fi


\begin{abstract}
In this paper we predict sea surface salinity (SSS) in the Arctic Ocean based on satellite measurements. SSS is a crucial indicator for ongoing changes in the Arctic Ocean and can offer important insights about climate change. We particularly focus on areas of water mistakenly flagged as ice by satellite algorithms. To remove bias in the retrieval of salinity near sea ice, the algorithms use conservative ice masks, which result in considerable loss of data. We aim to produce realistic SSS values for such regions to obtain more complete understanding about the SSS surface over the Arctic Ocean and benefit future applications that may require SSS measurements near edges of sea ice or coasts. We propose a class of scalable nonstationary processes that can handle large data from satellite products and complex geometries of the Arctic Ocean. Barrier overlap-removal acyclic directed graph GP (BORA-GP) constructs sparse directed acyclic graphs (DAGs) with neighbors conforming to barriers and boundaries, enabling characterization of dependence in constrained domains. The BORA-GP models produce more sensible SSS values in regions without satellite measurements and show improved performance in various constrained domains in simulation studies compared to state-of-the-art alternatives. An \texttt{R} package is available \repourl. 
\end{abstract}


\spacingset{1.45}
\section{Introduction} \label{sec:intro}

The Arctic Ocean, the world's coldest ocean, is experiencing severe changes including the loss of sea ice \citep{cavalieri_arctic_2012}, the flux of freshwater \citep{haine_arctic_2015}, and water temperature rise \citep{polyakov_warming_2012} due to climate change. Anticipated impacts of these changes on multiple species and communities around the globe include sea-level rise, harm to marine habitats, changes in animal species' diversity, and increased risk of storms in coastal areas \citep{hassol_impacts_2004}. Sea surface salinity (SSS) is one of the key characteristics that reflect these changes as salinity levels respond to sea ice coverage, river discharge, and water cycle processes such as precipitation and evaporation \citep{fournier_sea_2020}. Because even small changes in SSS can yield dramatic effects on the water cycle and ocean circulation, monitoring SSS will offer important insights about climate and help inform realistic plans for climate change adaptation.

Salinity measurements are obtained largely through satellite products. Some commonly used products are provided by the National Aeronautics and Space Administration (NASA) Aquarius and Soil Moisture Active Passive (SMAP) satellites and the European Space Agency's Soil Moisture and Ocean Salinity (SMOS) mission. Even though they vary in spatial and temporal resolutions, they are consistent with one another in the Arctic Ocean \citep{fournier_evaluation_2019}. In this paper we use SMAP Jet Propulsion Laboratory (JPL) measurements.

\begin{figure}
\centering
\includegraphics[scale = 0.65]{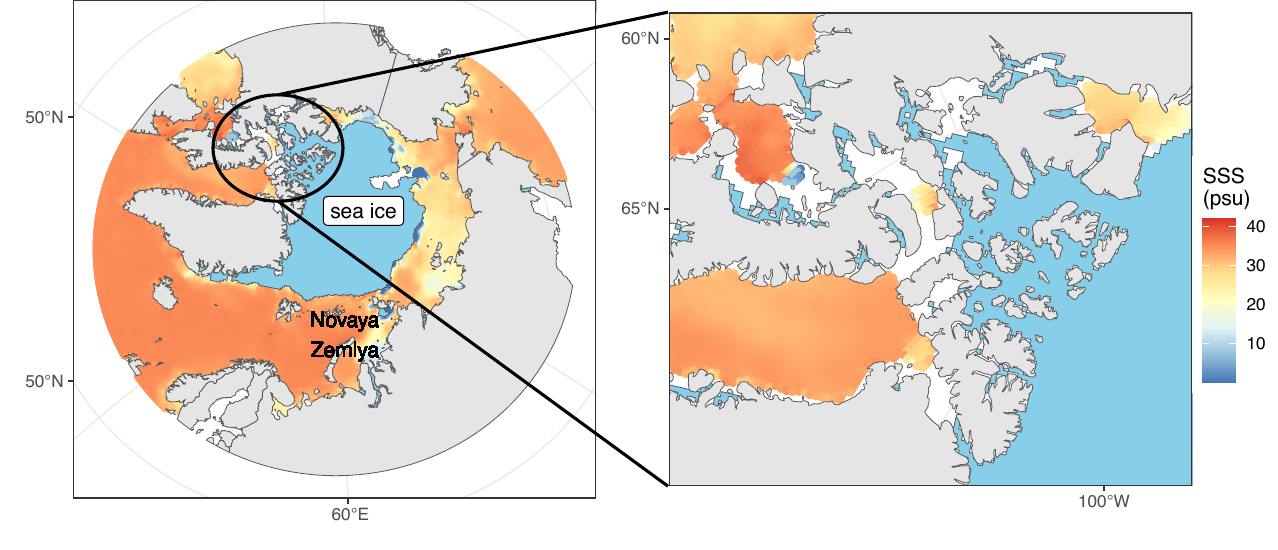}
\caption{The Arctic Ocean with SSS in practical salinity unit (psu) in August, 2020. Zoomed image of Canadian Arctic archipelago on the right panel. Transparent regions have no SSS measurements but are not part of sea ice.}
\label{fig:arctic_noobs}
\end{figure}

Although technical advances have increased accuracy of satellite products, some challenges remain. A major issue in current SSS retrieval algorithms is strict sea ice flagging \citep{meissner_smap_2021}. Since the salinity retrieval quality deteriorates near sea ice \citep{reul_sea_2020}, the algorithms use conservative ice masks to remove the effects of sea ice. Although SMAP JPL is known to use a more permissive ice mask than other satellite products \citep{fournier_evaluation_2019}, its algorithm is still subject to incorrect flagging of open water as sea ice and consequent removal of lots of data (see Figure \ref{fig:arctic_noobs}). We aim to develop a more complete spatial map of SSS in the Arctic Ocean by filling in the mismatched gaps between the actual ice extent and accidentally masked areas. This will improve our understanding on the distribution of SSS in missing regions and the connections between oceanic processes and sea ice changes \citep{tang_empirical_2021} and benefit other applications which require SSS measurements near edges of sea ice or coasts.

\begin{figure}
\centering
\subfloat[Average levels of daily SSS between July 20 and July 31, 2017.]{\includegraphics[scale = 0.43]{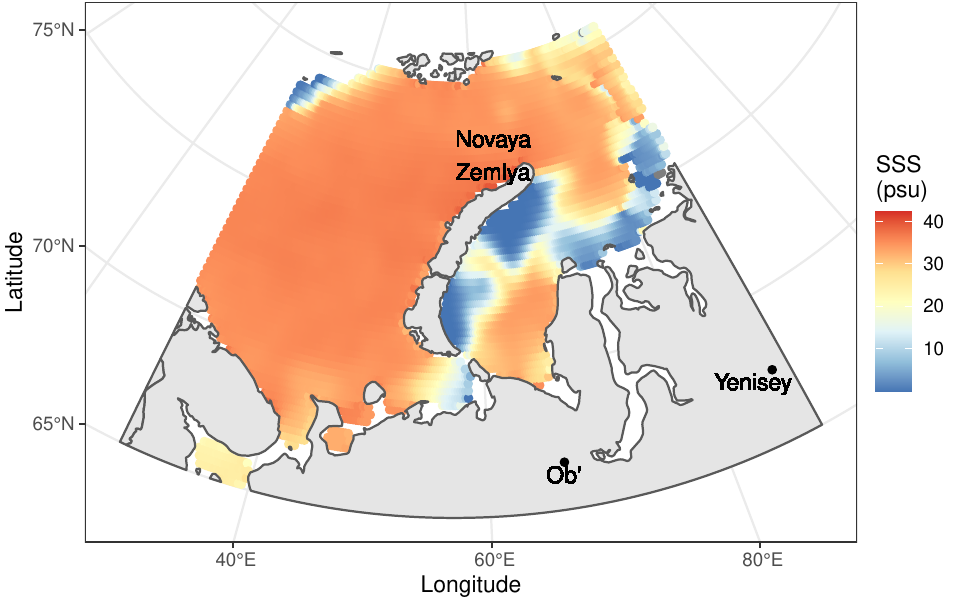} \label{fig:trueSSS}} \hspace{0.5em}
\subfloat[Monthly freshwater discharge by the Ob' and the Yenisey river averaged over 2017--2020.]{\includegraphics[scale = 0.43]{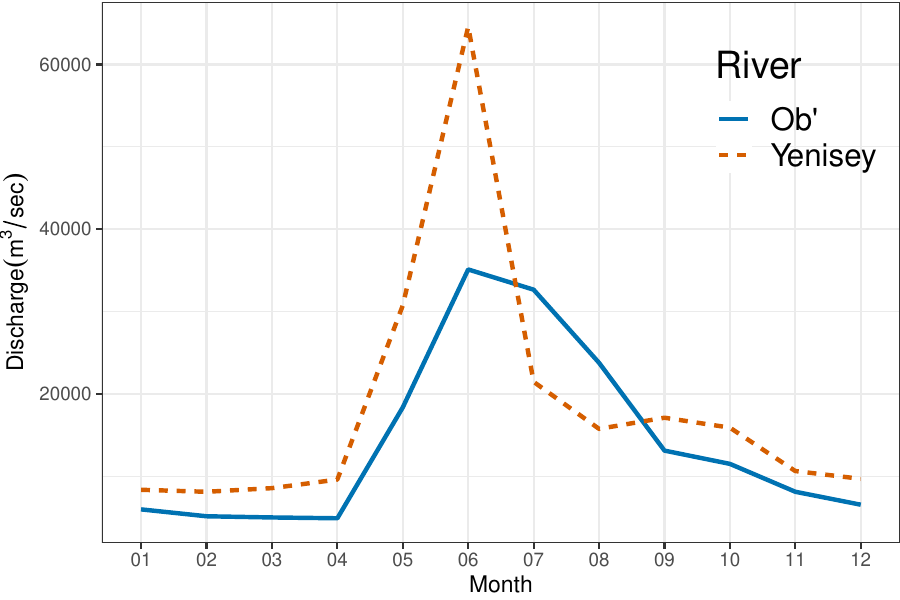} \label{fig:riverdischarge}}
\caption{Example and seasonal cause of nonstationary behavior of SSS in the Arctic Ocean.}
\label{fig:nonstmotif}
\end{figure}

For realistic and accurate SSS surface modeling, we need to handle three important aspects: (i) satellite products are large in size, (ii) the study region of interest, the Arctic Ocean, is a constrained domain involving complicated coastlines and peninsulas and islands of varying sizes, and (iii) such land masses create barriers which can significantly impact the distribution of SSS. In Figure \ref{fig:trueSSS} the average levels of daily SSS over late July in 2017 yield distinctive patterns blocked by Novaya Zemlya in the Arctic Ocean. Naturally, the correlation from a location $\bm{s}_0$ to another location $\bm{s}_1$ on the same side of the archipelago will be higher than that to $\bm{s}_2$ across the archipelago even if $\bm{s}_1$ and $\bm{s}_2$ have the same Euclidean distance away from $\bm{s}_0$. Typical stationary Gaussian process (GP) models, however, ignore the unique geometry of the domain and thus are likely to produce suboptimal results. In addition, they may not even be feasible due to poor scalability. Instead, the goal and the aspects of the application lead us to develop a stochastic approach that is scalable and characterizes nonstationary spatial correlations by boundaries and barriers. 

We are particularly motivated by GP models that achieve scalability by conditional independence assumptions based on sparse directed acyclic graphs (DAGs) \citep{vecchia_estimation_1988, stein_approximating_2004, emery_kriging_2009, gramacy_local_2015, datta_hierarchical_2016}. They impose well-structured sparsity in a precision matrix of a Gaussian likelihood by restricting neighbor sets from which each location receives directed edges. Various selection methods for neighbors have been proposed. Nearest-neighbor GP (NNGP; \citealt{datta_hierarchical_2016}) uses the closest locations in Euclidean distance, while \cite{stein_approximating_2004} advocate for also including some distant locations. Others suggest stepwise selection of the most relevant locations in terms of reducing kriging variance \citep{emery_kriging_2009} or predictive errors \citep{gramacy_local_2015}.

\begin{figure}
\centering
\includegraphics[scale = 0.65]{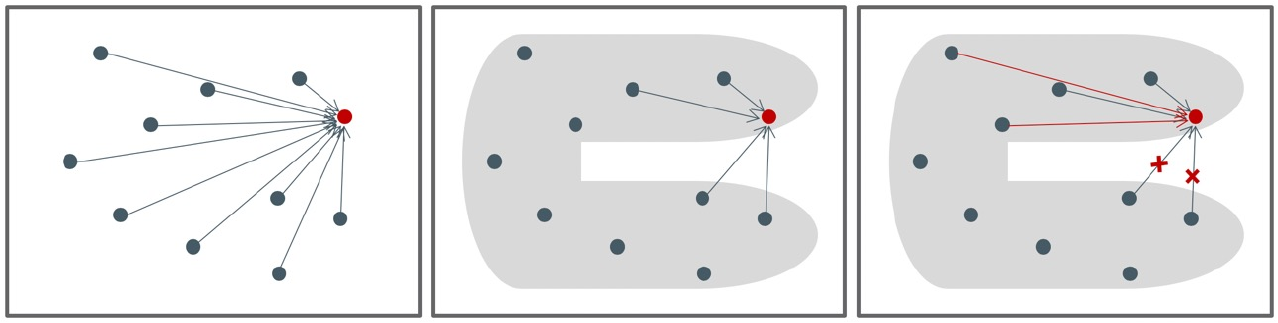}
\caption{Illustration of the traditional GP (left), NNGP (middle), and BORA-GP (right) using DAGs.}
\label{fig:DAGcartoon}
\end{figure}

We propose to construct neighbor sets conforming to barriers, which enables characterization of dependence in constrained domains. We give an illustrative example in Figure \ref{fig:DAGcartoon}. Given that the red point comes last in some ordering, the traditional GP assumes full dependence from the rest, inducing a directed edge from each point to the red one. NNGP reduces its neighbor set to the closest points, retaining directed edges from the four nearest locations if the number of neighbors is set to 4. However, when the domain is known to be the gray shaded area, one might reasonably doubt validity of the two neighbors below the red point because they are, in fact, far away from the red point given the intrinsic geometry of the domain. Therefore, we replace those by two other closest locations whose directed edge does not cross boundaries. We call this approach barrier overlap-removal acyclic directed graph GP (BORA-GP). We remove directed edges if they overlap barriers and construct a neighbor set with locations that are not only closest but also physically sensible and not blocked by barriers. By extending this geographically sensible DAG to the entire domain, our approach results in a scalable nonstationary process. 

To our knowledge, our approach is the first process-based attempt to predict SSS in regions incorrectly flagged as sea ice, while incorporating the complex coastlines of the Arctic Ocean. In the field of satellite SSS, efforts to correct for ice contamination have been devoted to technological advances \citep{kerr_present_2018} or empirical algorithms \citep{fore_combined_2016, meissner_smap_2021, tang_empirical_2021} during image reconstruction processing. The empirical algorithms aim to retrieve SSS as close to ice edges as possible and adjust for potential systematic biases in SSS near sea ice. They can produce near-ice SSS predictions that capture glacier melt and sharp salinity gradients away from the edges. However, these empirical approaches may lack universal applicability due to their specificity to a certain satellite. Furthermore, they show no statistically significant improvement in prediction biases when evaluated through cross-validation and comparisons with in situ data. Therefore, mitigation of the ice contamination effects is an ongoing focus of research \citep{reul_sea_2020}. 

The observation that current correcting algorithms do not lead to significant reductions in prediction biases may suggest that original SMAP data have sufficient accuracy and applicability to extend nearer to the ice edge in the Arctic Ocean without a need for additional bias correction, motivating our approach. Unlike aforementioned algorithmic approaches, BORA-GP can predict SSS right up to the edges of the sea ice, irrespective of ice concentration. BORA-GP can also incorporate spatial correlations and characterize uncertainty in predictions. 
    
\section{Relevant literature on constrained domains} \label{sec:litreview}

Previous literature on spatial smoothing over constrained domains typically involves partial differential equations (PDEs). Soap film smoother \citep{wood_soap_2008} and spatial regression with PDE (SR-PDE; \citealt{sangalli_spatial_2013}) are based on spline smoothing with PDE regularizations because PDEs can characterize complex phenomena, complying with boundaries and complicated geometries. \cite{bakka_non-stationary_2019} propose a Barrier Spatial Gaussian Field (Barrier SGF) with two stochastic PDEs (SPDEs), one for a domain and the other for barriers, whose solution is a GP with a \matern covariance function. In general, (S)PDE-based approaches require boundary conditions, such as fixing the value of a smooth function or its derivative along the boundary \citep{ramsay_spline_2002, lindgren_explicit_2011}. However, these conditions are often unrealistic, strongly affect the results, and produce anticonservative estimates of uncertainty near the boundary \citep{wood_soap_2008, bakka_non-stationary_2019}, leading \cite{sangalli_spatial_2013} to suggest mixed conditions along different partitions of the boundary and \cite{wood_soap_2008, bakka_non-stationary_2019} to propose more relaxed modifications assuming unknown boundary values. Moreover, analytic forms of solutions to (S)PDEs are unavailable. Thus, the above-mentioned methods resort to discrete mesh approximations over the domain, achieving computational efficiency of $\mc{O}(N^{3/2})$ at best, where $N$ is the number of nodes in the mesh and is not recommended to be much smaller than the number of observations $n$ \citep{sangalli_spatial_2021}.

Recently, intrinsic GPs (in-GPs; \citealt{niu_intrinsic_2019}) and Graph Laplacian based GPs (GL-GPs; \citealt{dunson_graph_2022}) have been proposed as GP approximations accounting for complex constrained domains. As a heat equation characterizes how a quantity like heat diffuses over a given complex domain, its solution, the heat kernel, is an appealing choice for GP covariance. However, the heat kernel is typically intractable, so it is approximated either utilizing Brownian motion \citep{niu_intrinsic_2019} or finitely many eigenpairs of the graph Laplacian \citep{dunson_graph_2022}. Unfortunately, the in-GP approximation is computationally very burdensome, as it depends on simulating Brownian motion multiple times. The GL-GP algorithm has the same order of computational complexity as usual GPs.

One could directly use a covariance function with geodesic distance, the distance of the shortest path between two locations, to explain intrinsic geometry of a domain. The geodesic distance can be useful to measure distances between locations in estuaries \citep{rathbun_spatial_1998} or vertices along the cortical surface of brain \citep{dai_bayesian_2021}. Although using geodesic distance is intuitive, the computational cost to estimate geodesic distance is not insignificant; state-of-the-art methods have computational complexity of $\mc{O}(N^2\log N)$ \citep{kirsanov_exact_2021} or $\mc{O}(n^2+\alpha n)$ \citep{li_geodesic_2020} with some constant $\alpha$. More critically, covariance functions with geodesic distance may be ill-defined. \cite{feragen_geodesic_2015} show that even a relatively simple covariance function $C(x,y) = \exp(-\lambda d(x,y)^q)$ is never positive-definite for all $\lambda >0$ if $q >2$ for any Riemannian manifold and its associated geodesic distance $d(x,y)$. The positive-definiteness of other covariance functions with geodesic distance remains largely unknown. In some cases approaches to approximate geodesic distance with Euclidean distance via a deformation (MDSdist; \citealt{loland_spatial_2003}) or to replace nonpositive eigenvalues of an invalid covariance matrix with a predetermined positive value (ClosePD; \citealt{davis_comparative_2019}) might be practical to ensure the positive-definiteness. However, the deformation approach fails to directly consider spatial dependence of the outcome of interest, and both approximation approaches are only feasible for small data ($\sim$400 observed locations and $\sim$11,000 prediction locations) due to their computational cost.

Our approach does not require boundary conditions or background knowledge about SSS values along coastlines and thus is free of these unrealistic assumptions. Furthermore, BORA-GP is a scalable and standalone process that can handle large satellite data and enable coherent interpolations of SSS at any new locations in the Arctic Ocean. Lastly, BORA-GP works with any Euclidean covariance functions without concerns for positive-definiteness. We employ BORA-GP as a sparsity-inducing prior for a latent spatial process in a Bayesian hierarchical framework.

\section{SSS and sea ice data in the Arctic Ocean} \label{sec:data}

We study SSS in the Arctic Ocean above 57$^{\circ}$N latitude using the SMAP product (V5.0) at Level 3 processed at NASA JPL. We use a monthly product at a 60 km spatial resolution and focus on August in 2020 because sea ice retreats the most during August to October in the Arctic Ocean. The satellite-derived SSS is as high as 42.124 practical salinity units (psu) and presented in Figure \ref{fig:arctic_noobs}. The number of SSS measurements in the study region is 50,225; for this data size, we cannot apply GPs without a scalable algorithm \citep{datta_nearest-neighbor_2016}. 

We chose SMAP JPL because its retrieval algorithm is thoroughly based on a variety of ancillary data. They include sea surface temperature (SST), wind speed and direction, land mask, precipitation, dielectric constant, galactic map, sea ice fraction, significant wave height and Hybrid Coordinate Ocean Model (HYCOM) for reference, where HYCOM utilizes sea surface currents, height, ocean turbidity, and heat flux along with other relevant predictors. SMAP JPL data are available at the NASA Physical Oceanography Distributed Active Archive Center (\url{https://podaac-tools.jpl.nasa.gov/las/UI.vm}). 

The monthly ice extent in the Arctic Ocean for August 2020 is obtained from the National Snow and Ice Data Center (NSIDC) Sea Ice Index Version 3 \citep{fetterer_sea_2017} available at \url{https://nsidc.org/data/G02135/versions/3}. Its superior quality as an Arctic sea ice extent indicator is confirmed through comparison with other leading indicators \citep{diebold_optimal_2021}. Taking it as ground truth of the Arctic sea ice, we predict missing data in the transparent regions in Figure \ref{fig:arctic_noobs} whose measurements were removed by the conservative retrieval algorithm of SMAP. At the same 60 km spatial resolution as SMAP, the missing data amount to 3702 locations that are more than 7\% of the total SMAP measurements in the Arctic Ocean. 

\section{BORA-GP} \label{sec:boragp}

Consider a univariate spatial process $\{w(\bm{s})\mid \bm{s}\in\mc{D}\subseteq \R^d\}$, where $d = 2$ is the dimension of a domain $\mc{D}$. We fix some ``reference'' locations in $\mc{D}$ and denote them by $\mc{R} = \{\bm{r}_1,\dots,\bm{r}_k\}$. Let $\mc{G} = \{A, E\}$ be a DAG with nodes $A = \{a_1,\dots,a_k\}$ and edges $E=\{([a_i] \rightarrow a_i)\mid i=1,\dots,k\}$, where $w(\bm{r}_i)$ is mapped to $a_i$, and $[a_i]$ is called a neighbor set of $a_i$. We then define a joint density of $\vt{w}_{\mc{R}}=(w(\bm{r}_1),\cdots,w(\bm{r}_k))^T$ by imposing conditional independence assumptions based on $\mc{G}$, 
\begin{align}
    \tilde{p}(\vt{w}_{\mc{R}}) = \prod_{i=1}^kp(w(\bm{r}_i)\mid\vt{w}_{[\bm{r}_i]}) \label{eq:ptilde_wR}
\end{align}
where $p$ is some base density, $[\bm{r}_i] \subset \{\bm{r}_1,\dots,\bm{r}_{i-1}\}$ is a neighbor set for $\bm{r}_i$ with some arbitrary ordering, and $\vt{w}_{[\bm{r}_i]}$ contains realizations of $w(\bm{s})$ over $[\bm{r}_i]$. The set $[\bm{r}_i]$ is of size $\min(m,i-1)$ with some fixed number $m$. For a ``nonreference'' location
$\bm{u} \in \mc{U}=\mc{D} \setminus \mc{R}$, we assume conditional independence given a neighbor set $[\bm{u}]\subset \mc{R}$ of size $m$, leading to the following conditional joint density:
\begin{align}
    \tilde{p}(\vt{w}_{\mc{U}}\mid\vt{w}_{\mc{R}}) = \prod_{\bm{u}\in\mc{U}}p(w(\bm{u})\mid\vt{w}_{[\bm{u}]}). \label{eq:ptilde_wU}
\end{align}
With equations \eqref{eq:ptilde_wR} and \eqref{eq:ptilde_wU}, we can define the joint density of $\vt{w}_{\mc{L}}$ for any collection of locations $\mc{L}\subset\mc{D}$, which is easily proved to satisfy Kolmogorov conditions \citep{datta_hierarchical_2016}, implying that we have derived a valid stochastic process associated with the density.

{\linespread{1.0}\selectfont
\begin{algorithm}[H]
\DontPrintSemicolon
  \KwInput{The number of neighbors $m$, reference set $\mc{R}=\{\bm{r}_1,\dots,\bm{r}_k\}$, and barriers $\mc{B}$ 
  }
  \KwOutput{Neighbor set $[\bm{r}_i]\subset \mc{R}$ of size $\min(m, i-1)$, $\forall\bm{r}_i\in\mc{R}$ }
  \KwNote{The reference locations are arranged in some ordering, which ensures the first $m$ locations not blocked by $\mc{B}$.}
  
  \tcc{First-order neighbors}
  $[\bm{r}_1] = \emptyset$ \\
  \For{$i=2,\dots,m+1$} {$[\bm{r}_i] = \{\bm{r}_1,\dots,\bm{r}_{i-1}\}$}
  \For{$i=m+2,\dots,k$ \label{alg:refnb_first}}{ 
      Sort\footnotemark[1] $\{\bm{r}_1,\dots,\bm{r}_{i-1}\}$ by Euclidean distance to $\bm{r}_i$, enumerated by $\{\bm{r}_{\pi_1},\dots,\bm{r}_{\pi_{i-1}}\}$. \\
      \For{$j=1,\dots,i-1$}{
        \If{${\overrightarrow{\bm{r}_{\pi_j}\bm{r}_i}}$ does not intersect with $\mc{B}$}{$\bm{r}_{\pi_j}\in [\bm{r}_i]$}
        \If{$\#([\bm{r}_i])$\footnotemark[2] $=m$}{\textbf{break}}
      }
  \label{alg:refnb_firstend}}
  
  \tcc{Second-order neighbors}
  \For{$i$ such that $0<\#([\bm{r}_i])=m_1<m$ \label{alg:refnb_second}}{
  Sort $[\bm{r}_i]_2 := [[\bm{r}_i]]$\footnotemark[3] $\setminus [\bm{r}_i]$ by difference between sum of Euclidean distances and direct Euclidean distance.\footnotemark[4]
  
  \If{$\#([\bm{r}_i]_2)\geq m-m_1$}{Include the first $m-m_1$ elements of $[\bm{r}_i]_2$ in $[\bm{r}_i]$.}
  \Else{Include all elements of $[\bm{r}_i]_2$ in $[\bm{r}_i]$ and repeat from step \ref{alg:refnb_second} as necessary.}
  \label{alg:refnb_secondend}}
\caption{Neighbor search for reference locations}
\label{alg:refnb}
\end{algorithm}
}

\footnotetext[1]{Always in ascending order.}
\footnotetext[2]{$\#([\bm{r}_i])$ is the number of elements in $[\bm{r}_i]$.}
\footnotetext[3]{$[[\bm{r}_i]]$ is the union of neighbor sets of all elements in $[\bm{r}_i]$.}
\footnotetext[4]{\{dist$(\bm{r}_i, \bm{r}_{i'})$ +  dist$(\bm{r}_{i'}, \bm{r}_{i''})$\} - dist$(\bm{r}_i, \bm{r}_{i''})$ given $\bm{r}_{i'}\in [\bm{r}_i]$ and $\bm{r}_{i''} \in [\bm{r}_{i'}] \subset [[\bm{r}_i]]$.}

There are various ways to specify a sparse DAG $\mc{G}$. If $[\bm{r}_i] = N(\bm{r}_i)$, where $N(\bm{r}_i)$ is at most $m$ nearest neighbors of $\bm{r}_i$ by Euclidean distance, then the derived process becomes the well-known nearest neighbor process \citep{datta_hierarchical_2016}. Alternatively, we let $[\bm{r}_i]$ conform to shape characteristics of the domain and be close to $\bm{r}_i$ by Euclidean distance, leading to BORA processes. We suggest neighbor searching algorithms provided in Algorithms \ref{alg:refnb} and \ref{alg:nonrefnb} to construct geographically sensible neighbor sets for reference locations and nonreference locations, respectively. We first search for first-order neighbors (steps \ref{alg:refnb_first}--\ref{alg:refnb_firstend} in Algorithm \ref{alg:refnb} and steps \ref{alg:nonrefnb_first}--\ref{alg:nonrefnb_firstend} in Algorithm \ref{alg:nonrefnb}). For any location $\bm{s} \in \mc{D}$, we repeat collecting a neighbor in $\mc{R}$ whose directed edge to $\bm{s}$ never crosses barriers $\mc{B}$ in ascending order by Euclidean distance and stop if the resulting neighbor set $[\bm{s}]$ has $m$ elements. The neighbor searching pool is previously ordered reference locations if $\bm{s}=\bm{r}_i\in\mc{R}$ for some $i$ and the whole $\mc{R}$ if $\bm{s}\notin \mc{R}$. In cases where the resulting neighbor set has fewer than $m$ elements, we consider neighbors of the first-order neighbors, denoted by second-order neighbors or $[\bm{s}]_2$ (steps \ref{alg:refnb_second}--\ref{alg:refnb_secondend} in Algorithm \ref{alg:refnb} and steps \ref{alg:nonrefnb_second}--\ref{alg:nonrefnb_secondend} in Algorithm \ref{alg:nonrefnb}). One may stop with the first-order neighbors of size less than $m$. However, we found that using less than $m$ neighbors negatively impacts prediction accuracy, and thus we further look for the second-order neighbors, if needed.

{\linespread{1.0}\selectfont
\begin{algorithm}[H]
\DontPrintSemicolon
  \KwInput{The number of neighbors $m$, reference set $\mc{R}=\{\bm{r}_1,\dots,\bm{r}_k\}$, and barriers $\mc{B}$ 
  }
  \KwOutput{Neighbor set $[\bm{u}]\subset \mc{R}$ of size $m$, $\forall\bm{u}\in\mc{U}$ }

  \tcc{First-order neighbors}
  \For{$\bm{u}\in\mc{U}$ \label{alg:nonrefnb_first}}{
      Sort $\mc{R}$ by Euclidean distance to $\bm{u}$, enumerated by $\{\bm{r}_{\pi_1},\dots,\bm{r}_{\pi_{k}}\}$. \\
      \For{$i=1,\dots,k$}{
        \If{${\overrightarrow{\bm{r}_{\pi_i}\bm{u}}}$ does not intersect with $\mc{B}$}{$\bm{r}_{\pi_i}\in [\bm{u}]$}
        \If{$\#([\bm{u}])=m$}{\textbf{break}}
      }
  \label{alg:nonrefnb_firstend}}
      
  \tcc{Second-order neighbors}
  \For{$\bm{u}$ such that $0<\#([\bm{u}])=m_1<m$ \label{alg:nonrefnb_second}}{
  Enumerate $[\bm{u}]$ as $\{\bm{r}_{\pi_{u,1}},\dots,\bm{r}_{\pi_{u,m_1}}\}$. \\
  Define $[\bm{r}_{\pi_{u,j}}]_{>} := \{\bm{r}_{i} \mid {\overrightarrow{\bm{r}_{i}\bm{r}_{\pi_{u,j}}}} \text{does not intersect with } \mc{B}~ \forall i>\pi_{u,j}\}$ for $j\in\{1,\dots,m_1\}$, and
  $[[\bm{u}]]_{>} := \cup_{j=1}^{m_1}[\bm{r}_{\pi_{u,j}}]_{>}$. \\
  Sort $[\bm{u}]_2 := ([[\bm{u}]] \cup [[\bm{u}]]_{>}) \setminus [\bm{u}]$ by difference between sum of Euclidean distances and direct Euclidean distance. \\ 
  Include the first $m-m_1$ elements of $[\bm{u}]_2$ in $[\bm{u}]$.
  \label{alg:nonrefnb_secondend}}
\caption{Neighbor search for nonreference locations}
\label{alg:nonrefnb}
\end{algorithm}
}

\begin{figure}
\centering
\subfloat[Example of the first-order (triangles) and the second-order (square) neighbors of the red point given the gray barrier.]{\includegraphics[scale = 0.57]{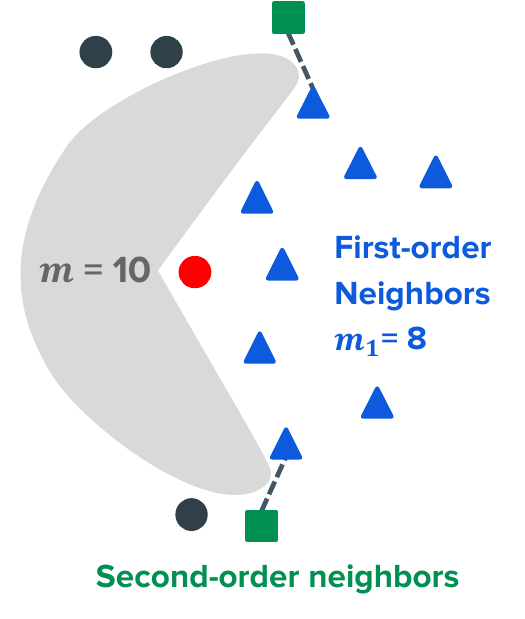} \label{fig:alg_12}} \hspace{0.2em}
\subfloat[Example of utilizing grid points (small circles) for the red point to escape from the isolated region and finding proxies of first-order (triangles) neighbors and their neighbors (second-order; square).]{\includegraphics[scale = 0.57]{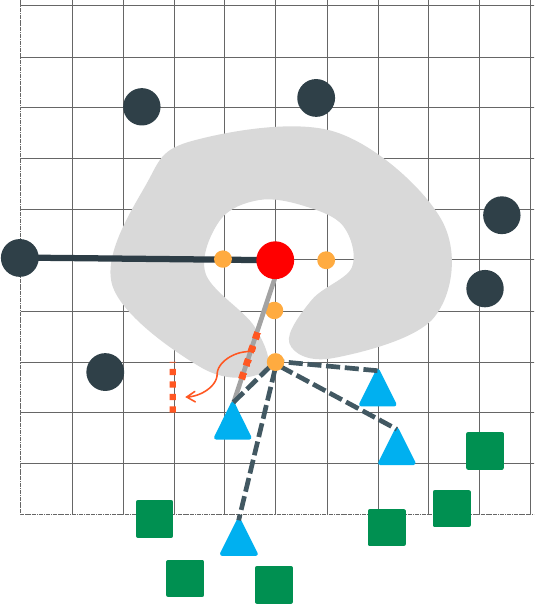} \label{fig:alg_grid}}
\caption{Visualization of the neighbor search algorithm.}
\label{fig:alg}
\end{figure}

This process is illustrated in Figure \ref{fig:alg_12} where the algorithm finds $m_1 = 8$ first-order neighbors and, among their neighbors conforming to a given domain, searches for two additional locations to obtain $m = 10$ neighbors in total. Clearly, all paths from $\bm{s}$ to $[\bm{s}]_2$ through the corresponding element in $[\bm{s}]$ do not cross barriers. We fill empty positions in the neighbor set with second-order neighbors in the order of smallest difference between sum of Euclidean distances through first-order neighbors and direct Euclidean distance. This promotes the selection of second-order neighbors reachable from $\bm{s}$ through a slight detour not deviating too much from a straight line. Neighbor selection for nonreference locations $\bm{u}$ should be irrelevant to ordering of $\mc{R}$. Therefore, $[\bm{u}]_2$ includes $[[\bm{u}]]_{>}$, the union of reference locations ordered later than each first-order neighbor and not crossing $\mc{B}$ so that any barrier noncrossing reference location can be selected as a neighbor regardless of ordering. 

With complex domains and certain choices of $\mc{R}$, some reference locations may have zero first-order neighbors. After the first-order neighbor search, if $[\bm{r}_i]=\emptyset$, then Algorithm \ref{alg:refnb_grid} in Supplementary Material \ref{suppsec:boragp} operates for $\bm{r}_i$ to escape from its isolated position. The algorithm creates an imaginary grid around the location where the grid is adaptive to the isolated region of a domain. Then it finds reference locations reachable with two straight lines through grid points, without being blocked by barriers, and accepts them as ``first-order neighbors.'' We return to step \ref{alg:refnb_second} in Algorithm \ref{alg:refnb} and proceed thereafter by including second-order neighbors until the number of neighbors reaches $m$. Figure \ref{fig:alg_grid} depicts these steps. For the red point without any first-order neighbor, our algorithm first finds a grid point that connects the red point to other reference locations outside the isolated region. We take those as proxies for first-order neighbors and fill a neighbor set of size $m$ with neighbors of the proxies. Detailed explanations about construction of a grid are provided in Supplementary Material \ref{suppsec:boragp}. 

We induce BORA-GP by allowing $\{w(\bm{s})\}$ to have $GP(0,\mbox{C}(\cdot, \cdot\mid \bm{\theta}))$ as a base process. A valid covariance function $\mbox{C}(\cdot, \cdot\mid \bm{\theta}):\mc{D}\times \mc{D} \rightarrow \R$ specifies associations across locations with parameters $\bm{\theta}$ such that $\mbox{C}(\bm{s}_i, \bm{s}_j\mid \bm{\theta}) = \mbox{cov}(w(\bm{s}_i), w(\bm{s}_j))$. This Gaussian assumption and equations \eqref{eq:ptilde_wR} and \eqref{eq:ptilde_wU} imply $\tilde{p}(\vt{w}_{\mc{R}}) = \prod_{i=1}^kN(w(\bm{r}_i);M_{\bm{r}_i}\vt{w}_{[\bm{r}_i]}, V_{\bm{r}_i})$ and $\tilde{p}(\vt{w}_{\mc{U}}\mid\vt{w}_{\mc{R}}) = \prod_{\bm{u}\in\mc{U}}N(w(\bm{u});M_{\bm{u}}\vt{w}_{[\bm{u}]}, V_{\bm{u}})$. For any location $\bm{s}\in\mc{D}$, $M_{\bm{s}}$ is defined as $C_{\bm{s},[\bm{s}]}C^{-1}_{[\bm{s}]}$ and $V_{\bm{s}}$ as $C_{\bm{s}} - C_{\bm{s},[\bm{s}]}C^{-1}_{[\bm{s}]}C_{[\bm{s}],\bm{s}}$, where $C_{\bm{s}}$ is the variance of $w(\bm{s})$ and $C_{\bm{s},[\bm{s}]}$ is the covariance matrix between $w(\bm{s})$ and $\vt{w}_{[\bm{s}]}$. Some algebra further simplifies $\tilde{p}(\vt{w}_{\mc{R}})$ to $N(\vt{0}, \tilde{C}_{\mc{R}})$, where $\tilde{C}^{-1}_{\mc{R}} = (I_k-M_{\mc{R}})^TV_{\mc{R}}^{-1}(I_k-M_{\mc{R}})$ with $I_k$ the $k$-dimensional identity matrix, $M_{\mc{R}}$ a $k\times k$ sparse matrix whose $i$th row has $M_{\bm{r}_i}$ over columns corresponding to $[\bm{r}_i]$ and zeros otherwise, and  $V_{\mc{R}}=\text{diag}(V_{\bm{r}_1},\cdots,V_{\bm{r}_k})$. The resulting BORA-GP$(0,\tilde{\mbox{C}}(\cdot,\cdot \mid \bm{\theta}))$ is a nonstationary GP with covariance function
\begin{align}
    \tilde{\mbox{C}}(\bm{s}_1,\bm{s}_2 \mid \bm{\theta}) = \left\{\begin{array}{ll}
         \tilde{C}_{\bm{r}_i, \bm{r}_j},& \text{if } \bm{s}_1=\bm{r}_i\in\mc{R} \text{ and } \bm{s}_2=\bm{r}_j\in\mc{R}  \\
         M_{\bm{s}_1}\tilde{C}_{[\bm{s}_1], \bm{r}_j}& \text{if } \bm{s}_1\notin\mc{R} \text{ and } \bm{s}_2=\bm{r}_j\in\mc{R} \\
         \bm{1}(\bm{s}_1=\bm{s}_2)V_{\bm{s}_1} + M_{\bm{s}_1}\tilde{C}_{[\bm{s}_1], [\bm{s}_2]}M_{\bm{s}_2}^T& \text{if } \bm{s}_1,\bm{s}_2\notin\mc{R}
    \end{array} \right. \label{eq:nonstcov}
\end{align} for any two locations $\bm{s}_1, \bm{s}_2 \in \mc{D}$ where $\bm{1}(\cdot)$ is the indicator function and $\tilde{C}_{P,Q}$ is a submatrix of $\tilde{C}_{\mc{R}}$ corresponding to locations in sets $P$ and $Q$. The covariance function is nonstationary because covariance between any two locations depends on $\mc{R}$ through their neighbor sets.

\begin{figure}
\centering
\includegraphics[scale = 0.7]{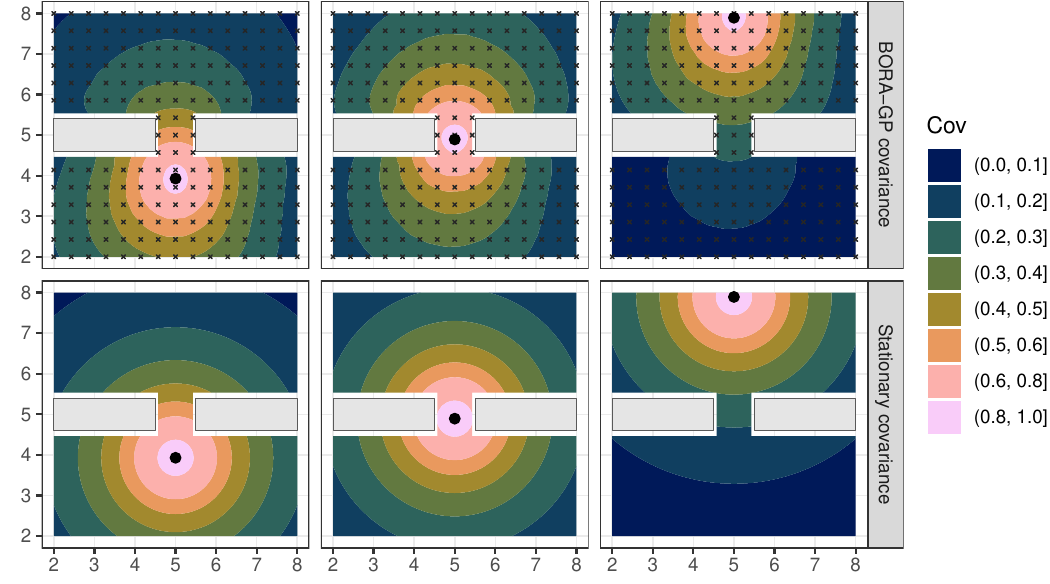}
\caption{Nonstationary covariance by BORA-GP (top) and the stationary covariance (bottom) between a dot and other locations in a domain with a slight opening between sliding doors. A grid marked with x is the reference set used to compute the BORA-GP covariance.}
\label{fig:covsliding}
\end{figure}

We empirically investigate behaviors of the resulting covariance function in equation \eqref{eq:nonstcov} in a domain $\mc{D}=[2,8]^2$ with barriers. The barriers are two sliding doors placed in the middle of $\mc{D}$, and they have an in-between opening of 1 unit. A 57$\times$57 grid makes all locations, and every fourth grid point populates the reference set of size $k=189$. The reference locations are marked with x in Figure \ref{fig:covsliding}. The locations are sorted in ascending order by $y$-coordinates. An exponential covariance function $\mbox{C}(\bm{s}_i,\bm{s}_j) = \sigma^2\exp(-\phi\|\bm{s}_i-\bm{s}_j\|)$ is used as a base covariance with $\sigma^2 = 1$ and $\phi=0.5$.

Figure \ref{fig:covsliding} compares the nonstationary covariance by BORA-GP with $m=10$ on the top row and the stationary exponential covariance on the bottom row at three different locations. Given the barriers, the nonstationary covariance by BORA-GP seems more reasonable. BORA-GP's nonstationary covariance is evidently pressed from both sides to squeeze in the small opening between the doors, while the stationary covariance ignores the doors and retains the perfect circle shape. More discussions about the nonstationary covariance by BORA-GP with different choices of $\mc{R}$ are provided in Supplementary Material \ref{suppsec:robust_to_R}.

Since nearest neighbor processes and BORA processes differ only in the construction of the edge set of $\mc{G}$, certain advantages of nearest neighbor processes apply directly. These include scalability, validity as a process bringing inferential advantages in estimation and prediction \citep{datta_hierarchical_2016}, and straightforward extensions to spatiotemporal \citep{datta_nonseparable_2016}, multivariate \citep{zhang_spatial_2022}, and non-Gaussian data \citep{peruzzi_spatial_2022}. Moreover, BORA-GP can enjoy efficient computational algorithms for posterior sampling \citep{finley_efficient_2019}. Refer to Section 3 of \cite{datta_hierarchical_2016} for detailed Bayesian implementation. 

Despite similarities, we emphasize that our ultimate goal is different from that of NNGP. We aim to cheaply derive nonstationary processes conforming to constrained domains from relatively simple base processes via a geographically sensible and sparse DAG; we do not intend to approximate the base process with the resulting process. The neighbor search algorithm and Markov chain Monte Carlo (MCMC) sampler for BORA-GP are implemented in the \texttt{R} package \texttt{boraGP}. The code to reproduce all analyses in this paper is provided \repourlanal. 

\section{Model comparisons} \label{sec:simulations}

This section compares BORA-GP with competitors for spatial random effects $\{w(\bm{s})\}$ with a response $\{y(\bm{s})\}$. NNGP is one of the most popular scalable approaches, to which BORA-GP reduces without barriers. Whenever feasible, we also fit barrier methods for constrained domains: Barrier SGF by the \texttt{R-INLA} package \citep{rue_approximate_2009}, Soap film smoother by the \texttt{mgcv} package \citep{wood_fast_2011}, SR-PDE by the \texttt{fdaPDE} package \citep{lila_fdapde_2020}, and MDSdist and ClosePD using code provided by authors of ClosePD. 

For methods that require a choice of covariance function, including BORA-GP, we use the \matern
\begin{align}
    \mbox{C}(\bm{s}_i,\bm{s}_j) = \frac{\sigma^2}{2^{\nu-1}\Gamma(\nu)}(\phi\|\bm{s}_i-\bm{s}_j\|)^{\nu}\mc{K}_{\nu}(\phi\|\bm{s}_i-\bm{s}_j\|), \label{eq:materncov}
\end{align}
where $\|\bm{s}_i-\bm{s}_j\|$ is the Euclidean distance between $\bm{s}_i$ and $\bm{s}_j$, $\nu > 0$ is a smoothness parameter, $\phi > 0$ controls spatial decay, and $\sigma^2$ is the partial sill representing spatial variability. $\mc{K}_{\nu}$ is a modified Bessel function of the second kind with order $\nu$. Our focus is on conducting accurate predictions in spatially constrained domains. We compare results in terms of Root Mean Square Prediction Error (RMSPE), Mean Absolute Prediction Error (MAPE), and empirical coverage probability and width of 95\% posterior predictive credible intervals. We denote posterior predictive credible intervals or confidence intervals by CI. For BORA-GP and NNGP, a set of observations plays the role of $\mc{R}$. Similarly, we use observations as knots in Soap film smoother. The smoothing parameter in SR-PDE is selected using cross-validation.

\subsection{Novaya Zemlya} \label{sec:novaya}

We first compare predictive performance of the different methods using SMAP data in a smaller region of the Arctic Ocean over a different time period. We use a random subsample of locations to predict the SSS values at held-out locations. We use SSS retrieved from daily SMAP production near Novaya Zemlya averaged over July 20--31 in 2017, yielding 8418 locations (see Figure \ref{fig:trueSSS}). We observe a dramatically diluted sea surface on the right-hand side of the archipelago within the Kara Sea. This reflects a continuation of the freshening phenomena around the Kara Sea during warm seasons observed in 2015 and 2016 \citep{tang_potential_2018}. This freshening is largely attributable to freshwater fluxes from two Arctic rivers, the Ob' and the Yenisey \citep{makinen_hydrographic_2016}. Figure \ref{fig:riverdischarge} exhibits a seasonal trend of the two rivers between 2017 and 2020 releasing large volumes of freshwater into the Kara Sea from May to September. The discharged freshwater accumulates, is carried over by currents and winds known to blow toward the north or west between July and October \citep{shestakova_climatology_2020}, and is blocked by Novaya Zemlya. Arctic rivers discharge data (Version 20220630) were obtained from Arctic Great Rivers Observatory at \url{https://arcticgreatrivers.org/discharge/}. 

We selected the size of the training data to mimic the sparsity of in situ SSS measurements. Although satellite data have much better sampling coverage and resolution, in situ data are used to validate satellite measurements subject to multiple disturbance factors \citep{reynolds_impacts_2005}. Argo is the primary source of in situ data worldwide and recorded 3306 salinity measurements in the Arctic Ocean in 2020 (\url{https://dataselection.euro-argo.eu/}). Over the same period, SMAP products at the 60 km spatial resolution result in 50,224 measurements, so that in situ data are $\sim7\%$ as large as SMAP. Based on this sparsity ratio, we subsampled 589 of 8418 satellite locations to serve as our training data, averaging results over 30 random training-test splits.

We assume $y(\bm{s}) = \beta_0 + w(\bm{s}) + \epsilon(\bm{s})$, where $y(\bm{s})$ is standardized SSS at location $\bm{s}$ with mean 0 and standard deviation 1, $\beta_0$ is an intercept, and $\epsilon(\bm{s}) \stackrel{iid}{\sim} N(0,\tau^2)$. We assess prediction accuracy at 7829 held-out locations, while also predicting at 2486 additional locations near Novaya Zemlya lacking satellite data. This data size exceeds the capability of MDSdist and ClosePD (see specifics in Section \ref{sec:litreview}); we obtained lack of memory errors in our implementation attempts. Furthermore, repeated attempts at using package \texttt{mgcv} to implement the Soap film smoother generated an error message of ``more knots than unique data values are not allowed;'' this error persists regardless of our knot specification, including for very small numbers of knots. In addition, in implementing SR-PDE using package \texttt{fdaPDE}, the \texttt{R} session automatically crashes upon execution of the function. We have implemented each of these competitors successfully in simulations. We conjecture that the above problems in attempting to implement the different packages for the real data application arise due to large data size and/or complexity of the domain. While it is possible that we could run the above competitors for subsampled data, we felt this was not worthwhile, as significantly downsampled data do not provide sufficient spatial resolution for adequate performance in this application. Hence, we focus on BORA-GP, NNGP, and Barrier SGF in analyzing the Novaya Zemlya data. 

The intercept $\beta_0$ is given a flat prior over $\R$, while $\tau^2\sim IG(2, 0.01)$ and $\sigma^2\sim IG(2, 1.269)$ so that $\sigma^2$ is loosely centered around the fitted value in a variogram analysis, with IG denoting an Inverse-Gamma distribution. A uniform prior is assigned to the decay parameter $\phi \sim Unif(0.002, 0.007)$ whose mean is the variogram estimate $\sim0.005$. Across methods, $\nu = 1$ is used in the covariance function in equation \eqref{eq:materncov} because Barrier SGF always fixes $\nu$ at 1. Penalized complexity (PC) priors \citep{simpson_penalising_2017} for parameters of Barrier SGF assume the range parameter $\sqrt{8}/\phi$ to be smaller than 1136 with prior probability 0.99, and the spatial standard deviation to be greater than 1 with prior probability 0.05. Results from BORA-GP and NNGP are based on 15,000 MCMC iterations after discarding 10,000 burn-in. 

\begin{figure}
\centering
\includegraphics[scale = 0.56]{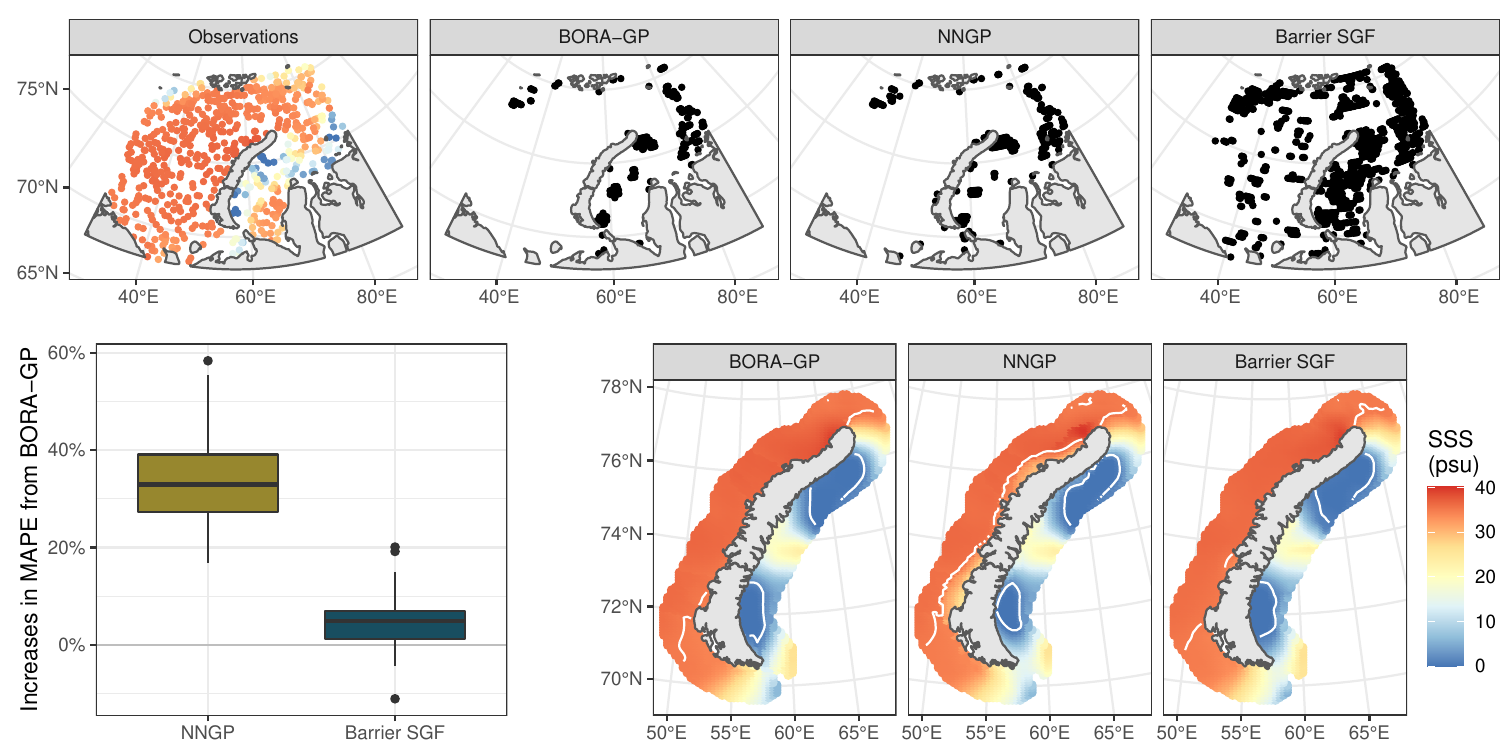}
\caption{Top panel shows locations at which 95\% predictive intervals specific to each method miss the true values, along with training data, from one randomly-selected replicate. The bottom-right panel illustrates predicted SSS bands around Novaya Zemlya at a higher resolution than the original SMAP resolution from another replicate. White lines are contour lines at 1 and 35 psu. The bottom-left panel displays how much NNGP and Barrier SGF increase MAPE relative to BORA-GP in the band around Novaya Zemlya across 30 replicates in percentage.}
\label{fig:novaya_pred}
\end{figure}

Figure \ref{fig:novaya_pred} demonstrates prediction improvements by BORA-GP, compared to two alternatives, which becomes more apparent in a band around Novaya Zemlya. Contour lines on the bottom-right panel show that SSS is underestimated on the left side and overestimated on the right side of Novaya Zemlya by NNGP. In contrast, BORA-GP and Barrier SGF successfully confine dilution to the Kara Sea, resulting in a stronger resemblance to the true surface in Figure \ref{fig:trueSSS}. Summaries of prediction performance metrics are provided in Table \ref{supptab:novaya_resall} in Supplementary Material \ref{suppsec:novaya}. The average RMSPE across replicates is lowest for BORA-GP (0.173), followed by Barrier SGF (0.179) and NNGP (0.185). The bottom-left panel of Figure \ref{fig:novaya_pred} shows BORA-GP significantly outperforms the competitors in terms of smaller prediction errors. NNGP increases MAPE relative to BORA-GP by at most 58\% and 35\% on average along the band around Novaya Zemlya. Barrier SGF increases MAPE as well, though to a lesser extent, by at most 20\% and 5\% on average compared to BORA-GP. RMSPEs convey similar results and are omitted to save space. 

Moreover, BORA-GP shows better performance with respect to the combination of out-of-sample coverage and width of 95\% predictive intervals. The average width of 95\% CIs is smaller for Barrier SGF (0.456) than for BORA-GP (0.701) or NNGP (0.723). However, the small width comes at the cost of under-coverage, with 95\% intervals having 0.882 coverage probability on average for Barrier SGF. Both BORA-GP and NNGP attain $\sim0.960$ average empirical coverage, with BORA-GP having narrower intervals on average. The top row of Figure \ref{fig:novaya_pred} shows that Barrier SGF misses true values much more than expected, particularly in the Kara Sea where decreased salinity is observed due to the freshwater fluxes. Inability to properly capture the effects of river discharge in SSS may lead to inaccurate assessment of the rate of warming on the Earth's surface. Increases in SSS draw more ocean heat into the deeper ocean and reduce the rate of warming. The freshwater fluxes, however, weaken this positive feedback loop and thus require special attention. Indeed, the Kara Sea is one of the regions with a vulnerable feedback loop. For example, \cite{nghiem_effects_2014} demonstrate that warm water discharges from rivers can accelerate the melting of sea ice whose broken pieces release a pulse of freshwater across the Arctic Ocean. Consequently, the sea surface is diluted even more, which expedites the warming. Therefore, identification of regions with lower SSS with proper uncertainty quantification is crucial to take appropriate actions before such regions fall into a vicious cycle of warming.

\begin{figure}
\centering
\includegraphics[scale = 0.55]{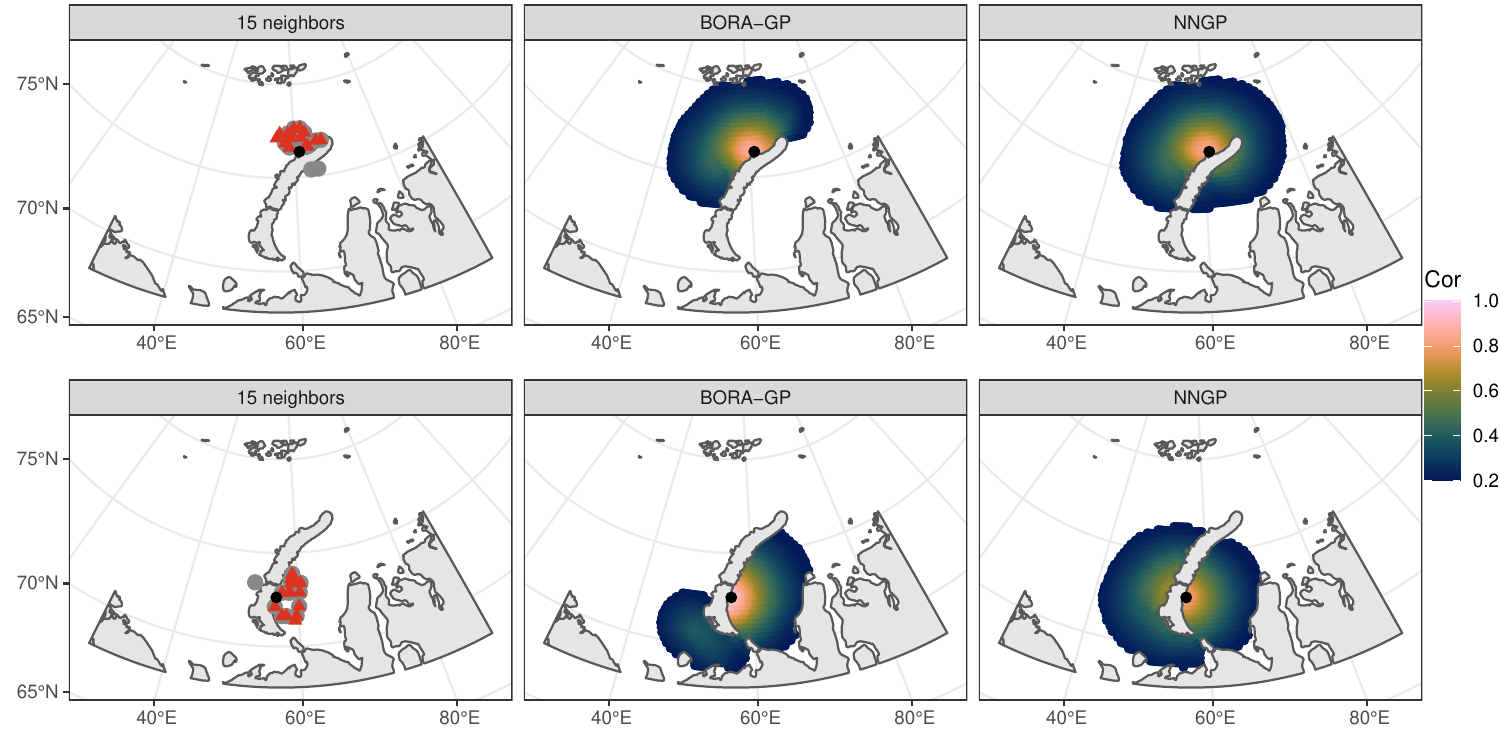}
\caption{Two examples (top row and bottom row) showing differences in neighbor selections and correlations by BORA-GP and NNGP. Triangles represent BORA-GP neighbors, and circles represent NNGP neighbors. Correlations larger than 0.2 are drawn for visualization purpose.}
\label{fig:novaya_cov}
\end{figure}

We further examine the neighbor selection and associated correlations of randomly selected locations by BORA-GP compared with NNGP. The number of neighbors $m$ is 15, and reference locations are sorted in ascending order by $x$-coordinates. Each row of Figure \ref{fig:novaya_cov} presents respective neighbors of a black point at around (77$^{\circ}$N, 60$^{\circ}$E) (top row) and (73$^{\circ}$N, 57$^{\circ}$E) (bottom row) and its correlations to the other locations. On the left facet of Figure \ref{fig:novaya_cov}, we observe that BORA-GP neighbors in triangles are more sensible, given the barriers. Some of NNGP neighbors cross barriers and are far away in terms of shortest path through the water. Other facets in Figure \ref{fig:novaya_cov} illustrate differences in the covariance between BORA-GP and NNGP by the different selection of neighbors; NNGP correlation ignores Novaya Zemlya while BORA-GP correlation flows along the archipelago. In regions without barriers, correlations by BORA-GP and NNGP display similar patterns. 

\subsection{Modified horseshoe} \label{sec:horseshoe}

\begin{figure}
\centering
\includegraphics[scale = 0.5]{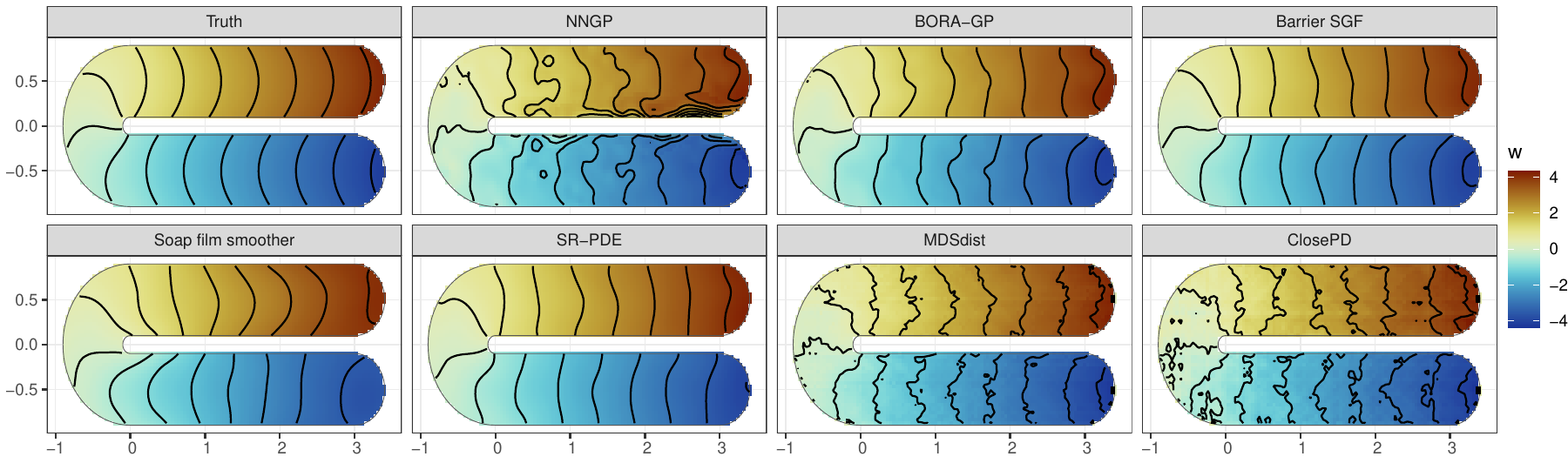}
\caption{True and predicted $w(\bm{s})$ surfaces from one (of 30) replicate with $n=300$ and $m=20$.}
\label{fig:horseshoe}
\end{figure}

We conduct a simulation study on the modified horseshoe example of \cite{wood_soap_2008}. The true surface of $w(\bm{s})$ on a fine grid of step size 0.04 is illustrated in Figure \ref{fig:horseshoe}. We obtain $y(\bm{s})$ by adding Gaussian noise $\epsilon(\bm{s}) \stackrel{iid}{\sim} N(0,\tau^2)$ to $w(\bm{s})$. The nugget is $\tau^2 = 0.01$. We vary the number of locations $n=300, 600, 1000$ in training data and the number of neighbors $m=10, 15, 20$ for BORA-GP and NNGP. For each combination of $n$ and $m$, $n$ locations are randomly selected out of 4046 total locations repeatedly for 30 replicates and sorted in ascending order by $x$-coordinates. Prior specifications are described in Supplementary Material \ref{suppsec:horseshoe}. Results for BORA-GP and NNGP are based on 5000 posterior samples after 5000 burn-in.

Figure \ref{fig:horseshoe} illustrates the true and predicted $w(\bm{s})$ surfaces from one randomly-selected replicate with $n=300$ and $m=20$. NNGP clearly misses the true contour pattern in the inner side of both arms (between -0.25 and 0.25 of the $y$-axis), whereas BORA-GP, Barrier SGF, Soap film smoother, and SR-PDE capture the bulging horseshoe well, even with relatively small $n$. Predictions from MDSdist and ClosePD include erroneous values on the boundary and highly noisy contour lines.

Full numerical results are available in Tables \ref{supptab:horseshoe_resall} -- \ref{supptab:horseshoe_resall_cont} in Supplementary Material \ref{suppsec:horseshoe}. The nugget $\tau^2$ is accurately estimated regardless of $n$ and $m$ by all methods except by MDSdist and ClosePD which estimate $\tau^2$ to be zero. According to the tables, all barrier methods but BORA-GP, Barrier SGF, and SR-PDE produce poorly calibrated predictions. MDSdist overestimates prediction variance, yielding wide 95\% CIs and inflated empirical coverage probability up to 0.999. In contrast, ClosePD suffers from low empirical coverage probability of 0.904 for $n=1000$. The issue of low empirical coverage probability also exists in Soap film smoother for any $n$ with at most 0.590 coverage for 95\% CIs. As Barrier SGF and SR-PDE exhibit better predictive performance amongst alternative barrier methods, we hereafter omit results from Soap film smoother, MDSdist, and ClosePD in this subsection.

\begin{figure}
\centering
\includegraphics[scale = 0.65]{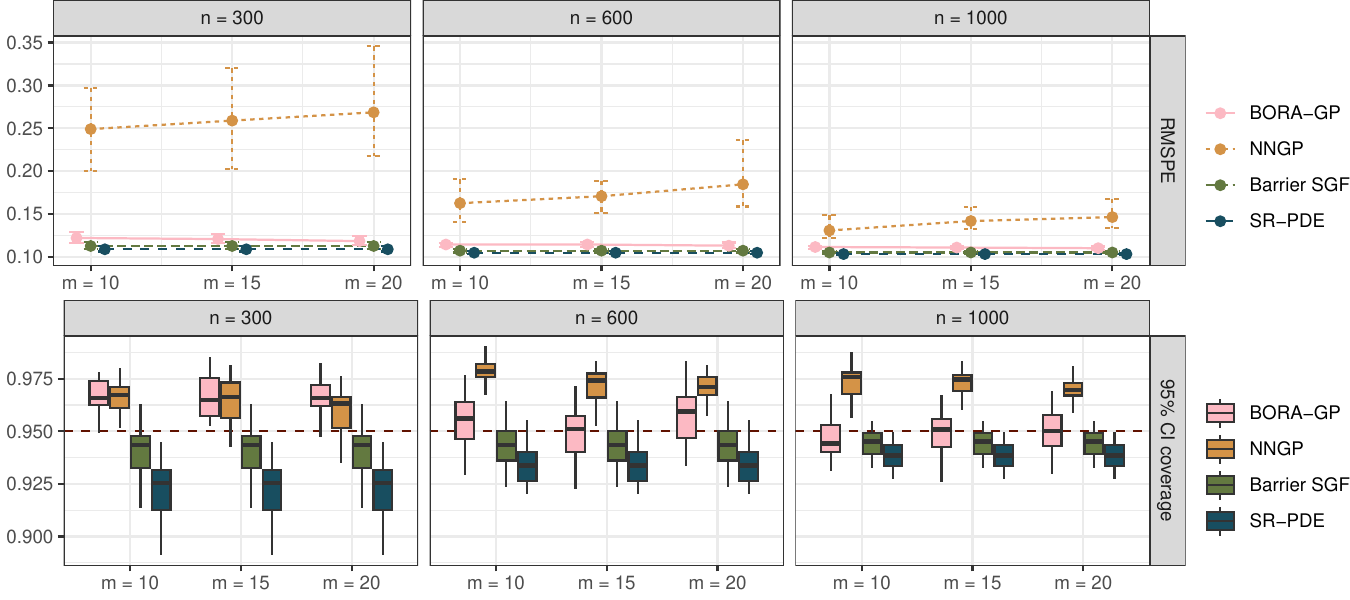}
\caption{Prediction performance for varying combinations of $n$ and $m$ in the modified horseshoe example. Top panel shows mean RMSPE as points and 0.025 and 0.975 quantiles as line ranges, and the bottom panel presents box plots of empirical coverage probability of 95\% CIs over 30 replicates at each $n$ and $m$. On the bottom panel, a dashed horizontal line is drawn at the nominal level 0.95 for comparison.}
\label{fig:horseshoe_pred}
\end{figure}

Figure \ref{fig:horseshoe_pred} presents RMSPE and 95\% CI coverage at each combination of $n$ and $m$. BORA-GP, Barrier SGF, and SR-PDE show RMSPEs of $\sim0.110$ for any $n$ with RMSPEs in SR-PDE slightly smaller than those in BORA-GP with a significant difference. NNGP produces substantially higher RMSPEs of $\sim0.260$ for $n = 300$ and $\sim0.140$ for $n = 1000$. Within each $n$ larger $m$ intuitively improves BORA-GP prediction and deteriorates NNGP prediction because the larger $m$ can mean more nonsensical neighbors in NNGP, given the domain. However, for BORA-GP the reduction in RMSPE by increasing $m$ diminishes for larger $n$. This suggests we may not lose much predictive power with modest values of $m (<20)$ for huge $n$, justifying our choice of $m = 15$ in Sections \ref{sec:novaya} \& \ref{sec:arctic}. Similar results are observed in terms of MAPEs available in Figure \ref{suppfig:horseshoe_pred} in Supplementary Material \ref{suppsec:horseshoe}. The number of neighbors $m$ affects not only prediction accuracy but also computation for both neighbor searching and model fitting in BORA-GP; smaller $m$ lowers the neighbor searching time and reduces the model fitting time (see Tables \ref{supptab:horseshoe_resall} -- \ref{supptab:horseshoe_resall_cont}).

In terms of empirical coverage of 95\% predictive intervals, all four methods show acceptable performance at varying degrees. BORA-GP and Barrier SGF produce 95\% CIs with associated empirical coverage probability close to 0.95 for $n \geq 600$, even though Figure \ref{fig:horseshoe_pred} shows that interquartile ranges of empirical coverage by Barrier SGF are always lower than the nominal level. SR-PDE also suffers low empirical coverage, which becomes worse for smaller $n$. NNGP yields 95\% CIs that slightly overcover, having average empirical coverage probability of $\sim0.970$. This suggests wider CIs by NNGP than by other methods, which is evident in summaries of 95\% CI width, as illustrated in Figure \ref{suppfig:horseshoe_pred} in Supplementary Material \ref{suppsec:horseshoe}. 

\subsection{Faults}

\begin{figure}
\centering
\includegraphics[scale = 0.45]{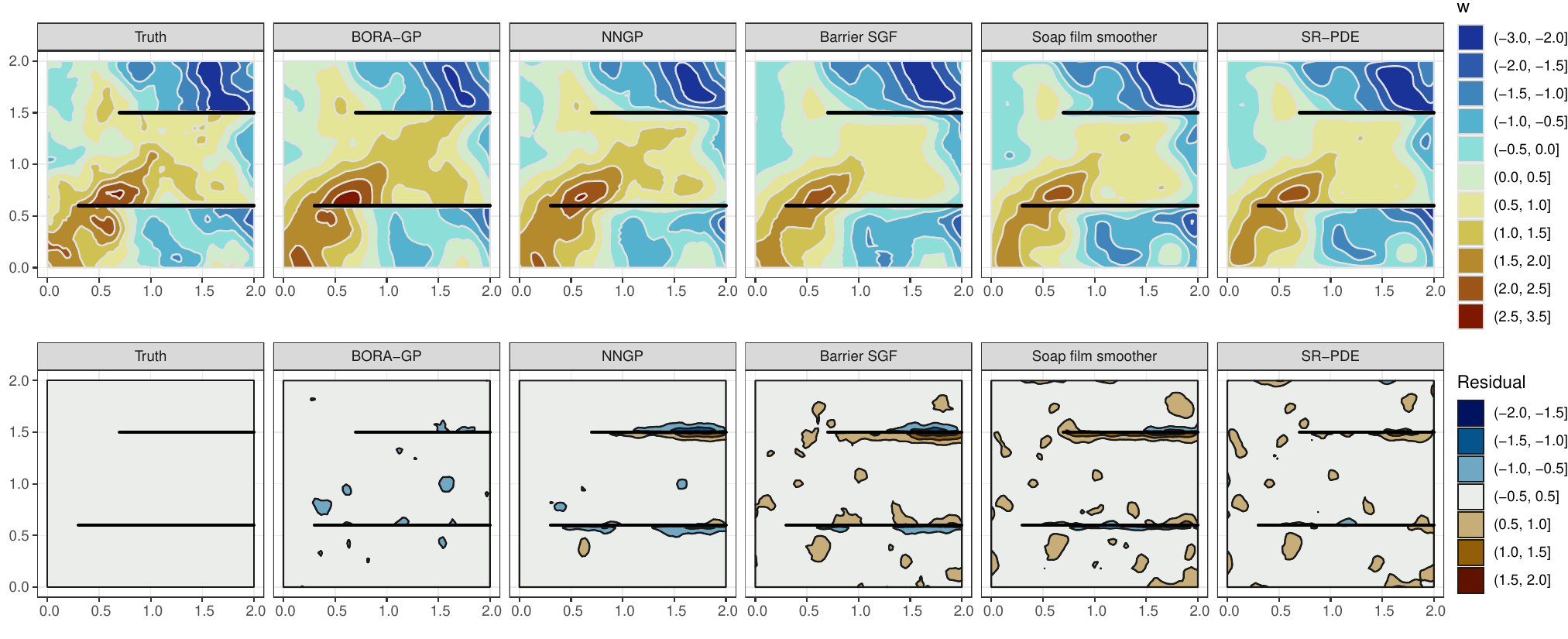}
\caption{True and predicted $w(\bm{s})$ surfaces and residual ($w(\bm{s})-\hat{w}(\bm{s})$) surfaces from one (of 30) synthetic data set.}
\label{fig:faults_w}
\end{figure}

BORA-GP is the true data generating process in this simulation study. Figure \ref{fig:faults_w} displays horizontal lines, two faults of length 1.7 and 1.3, on $\mc{D}=[0,2]^2$ in which 4489 locations are created on a $67\times 67$ grid. At each location $\bm{s}$, the regression model $y(\bm{s}) = \beta_0 + x(\bm{s})^T\beta_1 + w(\bm{s}) + \epsilon(\bm{s}),~\epsilon(\bm{s}) \stackrel{iid}{\sim}N(0,\tau^2)$ generates $y(\bm{s})$ with $x(\bm{s})\stackrel{iid}{\sim}N(0,1)$, $\beta_0=1$, $\beta_1=0.5$, and $\tau^2=0.1$. We assume BORA-GP with the base covariance function in equation \eqref{eq:materncov} for the true latent spatial effects $w(\bm{s})$ with $\nu=1.5$, $\sigma^2 = 1$, and $\phi=4$. The true number of neighbors is 15. We use $n=408$ observations as training data, which are fixed at a $12\times34$ grid and sorted in ascending order by $y$-coordinates. We vary $m$ to be 10, 15, and 20 for fitting BORA-GP and NNGP and save 5000 posterior samples after 10,000 burn-in. Supplementary Material \ref{suppsec:faults} describes detailed prior specifications. 

Complete results are provided in Tables \ref{supptab:faults_resall} -- \ref{supptab:faults_resall_cont} in Supplementary Material \ref{suppsec:faults}. Different choices of $m$ have minimal impact on BORA-GP and NNGP. The tables show that BORA-GP correctly recovers parameters, reduces prediction errors compared to other methods, and produces well-calibrated predictions that cover held-out empirical observations at the appropriate level. In BORA-GP with $m = 10$, the average of posterior means of model parameters over the replicates are 1.032 for $\beta_0 = 1$, 0.506 for $\beta_1 = 0.5$, 0.100 for $\tau^2 = 0.1$, 0.872 for $\sigma^2 = 1$ and 4.458 for $\phi = 4$. In addition, the average of the consistently estimable parameter $\sigma^2\phi^{2\nu}$ in a \matern covariance function \citep{zhang_inconsistent_2004} is $71.997$ with its standard error of $16.962$ for the true value 64. Overall, parameter estimates from BORA-GP are closer to the true values than estimates of the competitors. The average RMSPEs increase in the order of BORA-GP (0.368 with $m=20$), NNGP (0.393 with $m=20$), Barrier SGF (0.397), SR-PDE (0.412), Soap film smoother (0.414), MDSdist (0.460), and ClosePD (0.525), whose differences from BORA-GP are significant for the latter four methods. Poor average empirical coverage of 95\% CIs of $y(\bm{s})$ at unobserved locations $\bm{s}$ is observed in Soap film smoother (0.642), with ClosePD (0.921) also below the nominal level. 

Tables \ref{supptab:faults_resall} -- \ref{supptab:faults_resall_cont} present the computation time divided into preprocessing and model fitting. Neighbor search in BORA-GP, mesh construction in Barrier SGF, mesh and finite element basis creation in SR-PDE, or distance and covariance estimation in MDSdist/ClosePD counts as preprocessing which is done only once. Compared to the horseshoe domain in Section \ref{sec:horseshoe}, the current domain is more complex, yielding longer preprocessing time. The computational time of the model fitting step of BORA-GP does not depend on the complexity of the domain. 

Figure \ref{fig:faults_w} illustrates predicted surfaces of $w(\bm{s})$ with $m=15$ and corresponding residual surfaces in one synthetic data set. MDSdist and ClosePD are missing as they tend to produce the worst predictions in $y(\bm{s})$, as discussed above. In the remaining methods, namely, NNGP, Barrier SGF, Soap film smoother, and SR-PDE, there is clear oversmoothing across faults, while BORA-GP properly captures the disconnection by two faults present in the true surface;  see positive contour lines around (0.5, 0.5) and negative contours on the right of the domain. Separate and disconnected contour lines in the truth and BORA-GP are wrongly connected in NNGP, Barrier SGF, Soap film smoother, and SR-PDE. As a result, in residual surfaces all competitors show significantly nonzero residuals around the faults. In particular, Barrier SGF accounts for barriers by fixing the range parameter in barriers to be a fraction of the range in a free domain \citep{bakka_non-stationary_2019}. This approach might not successfully handle thin barriers, such as faults, as this simulation study suggests. With extra simulations we found that Barrier SGF shows comparable performance to BORA-GP with 20 times thicker faults; see Supplementary Material \ref{suppsec:faults} for more discussions. 

\section{Application to SSS in the Arctic Ocean} \label{sec:arctic}

\begin{figure}
\centering
\subfloat[Order of locations and a simple illustration of Transpolar Drift in the Arctic Ocean in arrows.]{\includegraphics[scale = 0.38]{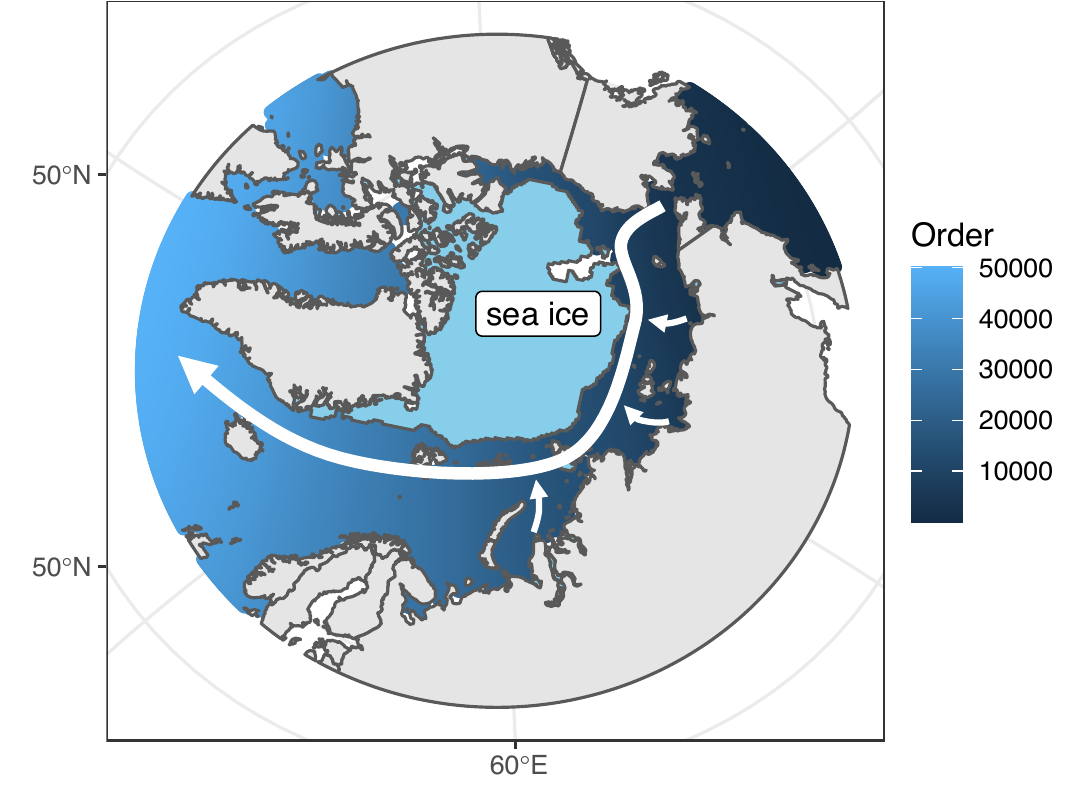} \label{fig:currents}} \hspace{0.2em}
\subfloat[Variogram of intercept-only model residuals.]{\includegraphics[scale = 0.43]{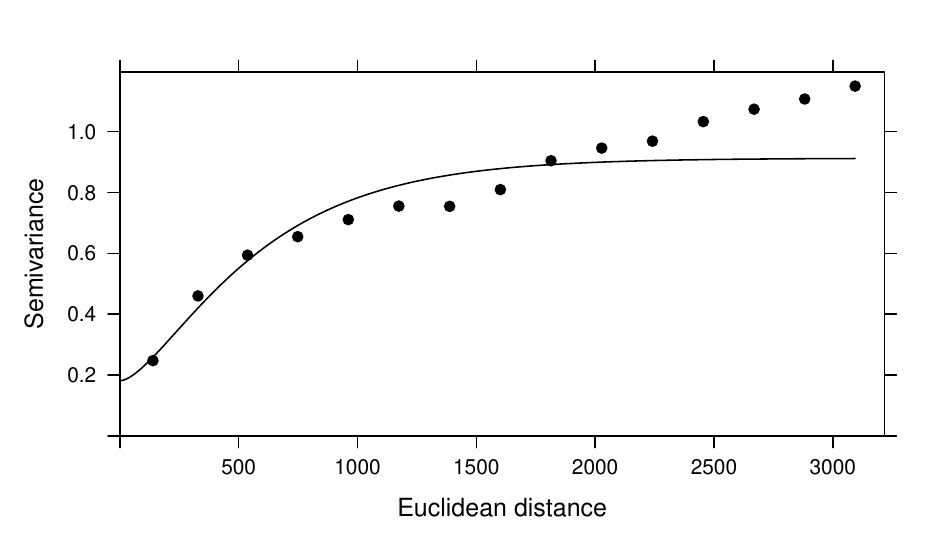} \label{fig:semivario}}
\caption{Model specifications based on the observed SSS data.}
\label{fig:emp}
\end{figure}

For each reference location, BORA-GP chooses $m = 15$ neighbors from previously ordered reference locations. Taking observations as $\mc{R}$, a sensible ordering would align the observed locations with ocean currents in this application. We focus on Transpolar Drift, a well-known major ocean current in the Arctic Ocean, drawn in arrows in Figure \ref{fig:currents} similarly to Figure 3 in \cite{carmack_toward_2015}. We arrange the locations in decreasing order of multiplication of $x$- and $y$-coordinates, which resembles Transpolar Drift as shown in Figure \ref{fig:currents}. Locations are darker early in the ordering where arrows start and become brighter later in the ordering along the path of the arrows. 

We have 50,225 observations and predict at 3702 new locations in the mismatched gaps between actual sea ice extent and SMAP's ice mask. Due to this large data size, MDSdist and ClosePD are not applicable. Possibly due to complexity of the domain, Soap film smoother suffered the same issue as described in Section \ref{sec:novaya}, and SR-PDE and Barrier SGF both failed to create a mesh due to ``internal errors.'' Hence, we fit BORA-GP and NNGP to obtain predicted surfaces of Arctic SSS and compare behaviors of GPs with and without taking into account complex geometry of the domain.

We let $y(\bm{s}) = \beta_0 + w(\bm{s}) + \epsilon(\bm{s})$ where $y(\bm{s})$ is the standardized SSS, $\epsilon(\bm{s}) \stackrel{iid}{\sim} N(0,\tau^2)$, and $w(\bm{s}) \sim \text{BORA-GP}(0,\tilde{\mbox{C}}(\cdot,\cdot \mid \bm{\theta}))$. A \matern covariance function in equation \eqref{eq:materncov} is selected as a base covariance function whose parameters are informed by the fitted variogram in Figure \ref{fig:semivario}. On intercept-only model residuals of $y(\bm{s})$, we fitted a \matern variogram model. The fitted variogram suggested $\sigma^2 = 0.792$, $\phi = 0.003$, $\tau^2 = 0.181$, and $\nu = 1$. The priors for the covariance parameters were specified to be distributed around these values: $\sigma^2 \sim IG(2, 0.792)$, $\phi \sim Unif(0.001, 0.004)$, $\tau^2 \sim IG(2, 0.181)$, and $\nu$ is fixed at 1. The intercept $\beta_0$ is assigned a flat prior over $\R$.

We base our analysis on 10,000 posterior samples obtained by running the MCMC sampler for 25,000 iterations, discarding 5000 burn-in, and saving every second iteration. The posterior mean of $\tau^2$ is 0.002. Convergence diagnostics in Supplementary Material \ref{suppsec:arctic} suggest adequate convergence and mixing. BORA-GP and NNGP predict negative SSS values at 89 and 85 locations, respectively, out of 3702 locations due to the Gaussian assumption of the residuals. Transforming the response, for example, through $\log y(\bm{s})$, led to worse predictive accuracy and the proportion of negative predicted values was only $\sim2\%$; hence, we applied a simple post hoc rounding up to zero approach. This can be viewed as an application of posterior projections, a well-established approach to include constraints in Bayesian models \citep{neelon_bayesian_2004, lin_bayesian_2014, zhang_bounded_2018}.

Before examining the predicted surfaces, we first validate our predictions. Prediction validation in this application is challenging because of scarcity of in situ SSS measurements in the Arctic Ocean, particularly near the ice edge in which we are most interested. Despite a limited focus on the coast of Greenland, we believe in situ salinity data from NASA's Ocean Melting Greenland (OMG) mission could serve our purpose. OMG collected oceanographic measurements from 2016 to 2021 to reveal a role of the ocean in Greenland's ice loss. We use one of the OMG's annual campaign data sets collected by Airborne eXpendable Conductivity Temperature Depth (AXCTD) instruments during August 24--31, 2020, which is somewhat aligned with the time frame of our SMAP data. OMG AXCTD salinity values measured at depth 0.72m are considered as in situ SSS. The data set is available at \url{https://doi.org/10.5067/OMGEV-AXCT1} and presented in Figure \ref{suppfig:omg} of Supplementary Material \ref{suppsec:arctic}. 

For each of 76 OMG AXCTD profiles, we found a prediction location closest to the profile in Euclidean distance. At those matching locations, we calculate absolute error of predicted values relative to OMG SSS. BORA-GP and NNGP have median absolute errors of 2.302 psu and 2.368 psu, respectively. Although this is suggestive of improvements by BORA-GP, the differences are not statistically significant. The improvements for BORA-GP should be larger across the Arctic Ocean because Greenland is surrounded by open water without barriers and with enough observed SMAP measurements to leverage on near the coast. For more complex coast lines, with peninsulas, islands, and other challenging features, the performance of NNGP will inevitably decline significantly relative to BORA-GP.

\begin{figure}[ht!]
\centering
\includegraphics[scale = 0.7]{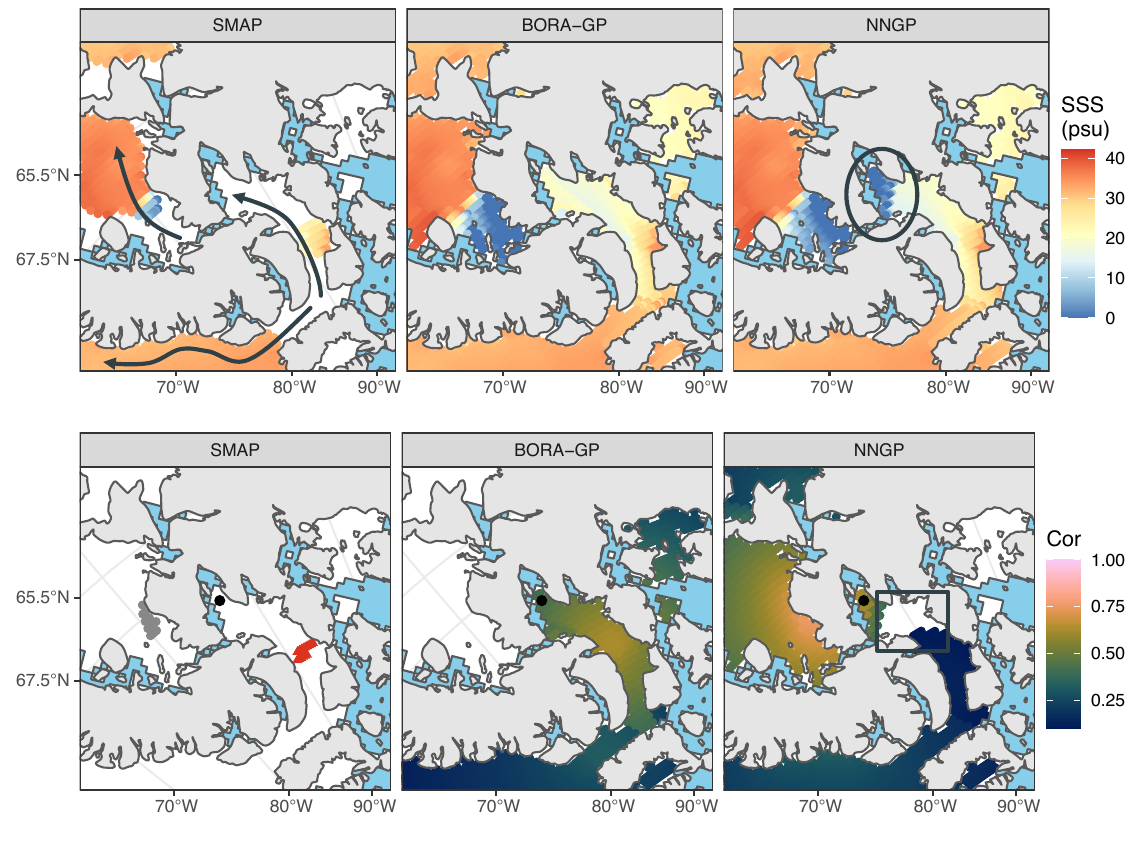}
\caption{The Canadian Arctic archipelago with SSS in August 2020. On the top row, missing regions under SMAP are filled with predicted SSS by BORA-GP and NNGP. The bottom row displays correlations larger than 0.1 from a black point at (67.6$^{\circ}$N, 87$^{\circ}$W) by the two methods. On the bottom-left panel, BORA-GP neighbors and NNGP neighbors for the black point are marked in red triangles and gray circles, respectively.}
\label{fig:arctic_pred}
\end{figure}

Although BORA-GP and NNGP produce similar predictions near the coast of Greenland, they show apparently distinctive behaviors in one of the missing regions in SMAP. Figure \ref{fig:arctic_pred} presents predicted surfaces of Arctic SSS on the top row and correlation surfaces on the bottom row in the Canadian Arctic archipelago. Slightly salty water on the right side of the archipelago and diluted water on the left side, known as the Northwestern Passages, are well separated by BORA-GP. Contrastingly, NNGP appears to let the diluted water on the left side of the archipelago flow toward the right side across the barrier, resulting in dilution in the circled area of Figure \ref{fig:arctic_pred}. This NNGP prediction may not be supported by the Arctic currents depicted as arrows on the first panel of Figure \ref{fig:arctic_pred}, drawn similarly to Figure 3.29 in \cite{amap_amap_1998}. According to the currents, saltier water on the right side of the archipelago is likely to flow toward the Northwestern Passages, not the other way. Furthermore, the spillover of diluted water in the Northwestern Passages toward the right side of the archipelago would have been restricted because a thin in-between opening was covered by ice in August 2020.

The bottom row of Figure \ref{fig:arctic_pred} explains where and how the differences in the predicted surfaces occur. In the facet of SMAP, 15 BORA-GP neighbors are compared with NNGP neighbors for the black point at (67.6$^{\circ}$N, 87$^{\circ}$W). All NNGP neighbors are selected from the Northwestern Passages despite highly restricted paths to connect them to the black point. In contrast, water can freely flow from BORA-GP neighbors toward the black point without barriers along the path of the known currents. These distinct sets of neighbors render drastically different correlation surfaces. BORA-GP correlations appear physically reasonable, whereas NNGP appears to completely flip the direction of the water flow. NNGP not only yields moderate to high correlations across barriers but also yields negligible correlations less than 0.1 at locations right next to the black point (see the rectangle in the NNGP facet). 

We place high emphasis on accurate prediction of SSS, particularly low values, because low salinity has great importance in the context of marine ecosystem as well as ocean processes and the rate of warming, as we already discussed above. For instance, low SSS is known to drive strong stratification of the ocean water, which slows the nutrient supply to the sea surface and limits phytoplankton productivity \citep{farmer_arctic_2021}. This strong stratification also affects the carbonate system of the ocean, leading to acidification \citep{polukhin_role_2019}. Ocean acidification has negative effects on marine species and marine food chains \citep{guinotte_ocean_2008} and weakens the role of reefs in projecting coastal areas against storms \citep{hoegh-guldberg_coral_2007}. Therefore, the inaccurate prediction of SSS by NNGP is undesirable, as it can hamper precise assessment of risks that human society and marine ecosystem are facing. Visual inspections of predicted SSS surfaces and correlation surfaces in relation to science, such as the Arctic currents, support that BORA-GP can bring inferential benefits in analyzing SSS of the Arctic Ocean, filling the mismatched gaps between actual sea ice extent and satellites' ice masks. 

\section{Discussion} \label{sec:discussion}

In this paper we introduced a scalable nonstationary spatial process model to predict Arctic SSS in missing regions mistakenly flagged as sea ice by the SMAP retrieval algorithm. To our knowledge, our model is the first process-based approach incorporating the complex geometry of the Arctic Ocean to complete SSS measurements. We first confirmed that our approach improved inferences and predictive accuracy over alternatives using a separate set of SSS data near Novaya Zemlya to understand the performance of BORA-GP in the context of the application. We then applied BORA-GP to gain extended understanding of SSS nearer to coasts and sea ice in the Arctic Ocean. We found that a method ignoring barriers is subject to implausible predictions and covariances. Alternative barrier methods were not fully scalable to massive data, needed configurational repairs for more complex domains, or suffered from poorly calibrated predictions, which in the Arctic Ocean may lead to incorrect assessment of ocean circulation, climate change, and threats to the marine ecosystem. We thoroughly investigated robustness of BORA-GP to reference sets and to their ordering in Supplementary Material \ref{suppsec:robust}.

There are several interesting extensions of our analysis. First, we may include covariates into regressions to increase prediction accuracy of SSS. We did not utilize any covariates in current regressions because relevant covariates, such as SST and precipitation, were mostly observed at lower resolutions than our prediction resolution. If any covariates appear important in explaining SSS and become densely available nearer to sea ice or coasts in the Arctic Ocean, they can be easily added to our analysis. Another extension is to accommodate non-Gaussian data. As climate change accelerates, it may become more common to observe extremely diluted sea surfaces, suggesting a need for flexible distributions. By incorporating heavier tailed parametric distributions or nonparametric distributions, these non-Gaussian SSS values could be directly fit without nonrealistic assumptions. We can also extend our approach for spatiotemporal data to analyze temporal as well as spatial changes of SSS in the Arctic Ocean. Furthermore, it would be natural to consider unknown or even changing barriers by time. Sea ice is one example of changing barriers. By extending our approach to spatiotemporal processes, which handle different barriers at each time point, we will be able to produce more dynamically changing salinity surfaces and make direct inferences on the relationship between SSS and varying sea ice coverage. 

Beyond the SSS application, BORA-GP may prove useful in other application contexts. Many applications in point-referenced geostatistics have measurements that are collected in constrained domains and show different patterns by physical barriers. Groundwater contamination in water bodies is an example as the water bodies usually have complex shapes and intricate interconnections to each other. Different wildlife landscapes or crime patterns in two neighborhoods, which are closely located but disconnected by highways or borders, provide other examples. In these applications BORA-GP will be better suited than other approaches by incorporating geometry of the domain. However, BORA-GP might be too restrictive in some applications because it employs the domain boundary as a hard criterion; the domain decides either the inclusion or exclusion of a location as a neighbor with probability 1. An interesting relaxation is to make it a soft boundary. For any location $\bm{s}\in\mc{D}$, we may assign different weights on reference locations. The weight of a reference location $\bm{r}$ decreases as either Euclidean distance between $\bm{s}$ and $\bm{r}$ increases or a directed edge from $\bm{r}$ to $\bm{s}$ overlaps barriers, ensuring neighbors crossing barriers can still receive a small but positive probability. We can also anticipate different applications may demand different assumptions about their boundaries: hard boundaries, soft boundaries, or a combination of the two.

\if0\blind
{
\section*{Acknowledgement} 
We are grateful for the financial support from the National Institute of Environmental Health Sciences through grants R01ES027498 and R01ES028804 and from the European Research Council under the European Union’s Horizon 2020 research and innovation programme (grant agreement No 856506). 
} \fi

\if0\standaloneappendix
{
\newpage
\begin{center}
{\spacingset{1} \LARGE\bf Supporting Materials for\\``\articletitle''}
\end{center}
\setcounter{section}{0}
\setcounter{equation}{0}
\setcounter{algocf}{0}
\setcounter{figure}{0}
\setcounter{table}{0}
\setcounter{page}{1}
\makeatletter
\renewcommand{\thesection}{S\arabic{section}}
\renewcommand{\theequation}{S\arabic{equation}}
\renewcommand{\thefigure}{S\arabic{figure}}
\renewcommand{\thetable}{S\arabic{table}}
\renewcommand{\thealgocf}{S\arabic{algocf}}

\section{BORA-GP} \label{suppsec:boragp}

After steps \ref{alg:refnb_first} -- \ref{alg:refnb_firstend} in Algorithm \ref{alg:refnb}, if $[\bm{r}_i]=\emptyset$ for some $i \in\{m+2, \dots, k\}$, we use Algorithm \ref{alg:refnb_grid} to find proxies of first-order neighbors before steps \ref{alg:refnb_second} -- \ref{alg:refnb_secondend} in Algorithm \ref{alg:refnb}. 

{\linespread{1.0}\selectfont 
\begin{algorithm}[H]
\DontPrintSemicolon
  \KwInput{The number of neighbors $m$, reference set $\mc{R}=\{\bm{r}_1,\dots,\bm{r}_k\}$, and barriers $\mc{B}$ 
  }
  \KwOutput{Non-empty neighbor set $[\bm{r}_i]\subset \mc{R}$ of at most size $m$}
  \KwNote{The reference locations are arranged in some ordering.}
  
  \For{$i \in \{m+2, \dots, k\}$ such that $[\bm{r}_i]=\emptyset$}{
      Among its $m$ nearest neighbors crossing $\mc{B}$, 
      find the longest Euclidean distance $r_l$ to $\bm{r}_i$ and the shortest overlapping length $d_l$ between $\mc{B}$ and the lines from $\bm{r}_i$ to the nearest neighbors.\\
      Create grid points with increments of $d_l$ in a square whose center is at $\bm{r}_i$ and the half length is $r_l$. Sort them by Euclidean distance to $\bm{r}_i$, enumerated by  $\{\bm{l}_1,\dots,\bm{l}_G\}$. \\
      \For{$t=1,\dots,G$}{
          \If{${\overrightarrow{\bm{l}_{it}\bm{r}_i}}$ does not intersect with $\mc{B}$}{
              \For{$j=1,\dots,m$}{
              \If{${\overrightarrow{\bm{r}_{\pi_j}\bm{l}_{it}}}$ does not intersect with $\mc{B}$}{$\bm{r}_{\pi_j}\footnotemark[1]\in [\bm{r}_i]$}
              }
          }
      \If{$\#([\bm{r}_i])>0$}{\textbf{break}}
      }
  }
  
\caption{Neighbor search for isolated reference locations}
\label{alg:refnb_grid}
\end{algorithm}
}

\footnotetext[1]{$\bm{r}_{\pi_j}$ as defined in Algorithm \ref{alg:refnb}}

The size and fineness of a grid square around each location vary in order to adapt to different isolated regions of a domain. In particular, our choice of a step size being the smallest overlapping length is to find the thinnest part of barriers around each location to the nearest neighbors. With such step sizes, we expect to find a balance between moving efficiently across and thoroughly around the barriers to find open spaces that could connect the isolated location and other reference locations. In Figure \ref{fig:alg_grid}, the shortest overlapping length $d_l$ is given in the orange dotted lines, and the half length $r_l$ of a grid square is in the dark solid line.

\section{Model comparisons}

\subsection{Novaya Zemlya} \label{suppsec:novaya}

\begin{table}[htbp]
    \centering
    \caption{Full simulation results on SSS data near Novaya Zemlya. Average (standard error) of prediction performance metrics and posterior mean of model parameters of interest, and average computation time are provided over 30 random training-test splits of SSS data near Novaya Zemlya.}
    \begin{tabular}{cccc}
    \hline
         & BORA-GP & NNGP & Barrier SGF\\ \hline 
   RMSPE & \textbf{0.173} (0.015) & 0.185 (0.016) & 0.179 (0.016) \\
   MAPE  & \textbf{0.071} (0.005) & 0.076 (0.006) & 0.072 (0.005) \\
   95\% CI coverage & 0.961 (0.010) & \textbf{0.959} (0.008) & 0.882 (0.044) \\
   95\% CI width & 0.701 (0.107) & 0.723 (0.086) & 0.456 (0.150) \\ 
   $\beta_0$ & -0.707 (0.099) & -0.553 (0.103) & -0.648 (0.376) \\ 
   $\tau^2$ & 0.006 (0.004) & 0.006 (0.003) & 0.004 (0.007) \\ 
   Preprocessing (sec.) & 115.365 & -- & 26.179 \\
   Runtime (sec.) & 284.480 & 286.280 & 28.954 \\ 
   Sec./iteration & 0.019 & 0.019 & -- \\ \hline
   \end{tabular}
   \label{supptab:novaya_resall}
\end{table}

Table \ref{supptab:novaya_resall} presents preprocessing time and total running time in seconds. Preprocessing indicates the neighbor search and barrier mesh construction for BORA-GP and Barrier SGF, respectively. The number of threads is set at 10 for model fitting. BORA-GP and NNGP samplers are based on MCMC, whose running time per iteration is additionally provided in seconds.

\subsection{Modified horseshoe} \label{suppsec:horseshoe}

BORA-GP and NNGP assign the same prior distributions for parameters: $\tau^2 \sim IG(2, 0.01)$, $\sigma^2 \sim IG(2, 6)$, and $\phi \sim Unif(1.230, 3.690)$. The uniform prior ensures that the spatial correlation can drop to 0.05 between 0.25 and 0.75 of the maximum distance of the horseshoe domain, and the $IG$ prior for $\sigma^2$ centers it loosely around the observed variability. Priors for parameters in Barrier SGF follow specifications in \cite{bakka_non-stationary_2019}. 

When implementing SR-PDE using package \texttt{fdaPDE}, an intercept is not allowed. Hence, we center the observed response to have mean zero before fitting SR-PDE and reinstate the sample mean for prediction. Preprocessing in Tables \ref{supptab:horseshoe_resall}--\ref{supptab:horseshoe_resall_cont} indicates the neighbor search, barrier mesh construction, mesh and finite element basis creation, or distance and covariance estimation for BORA-GP, Barrier SGF, SR-PDE, or MDSdist/ClosePD, respectively. Ten threads are used for model fitting, applicable only for BORA-GP, NNGP, and Barrier SGF. A large discrepancy in total runtime is partly attributable to differences in coding languages, which may complicate direct comparison between methods. 

\begin{figure}[!ht]
\centering
\includegraphics[width=\textwidth]{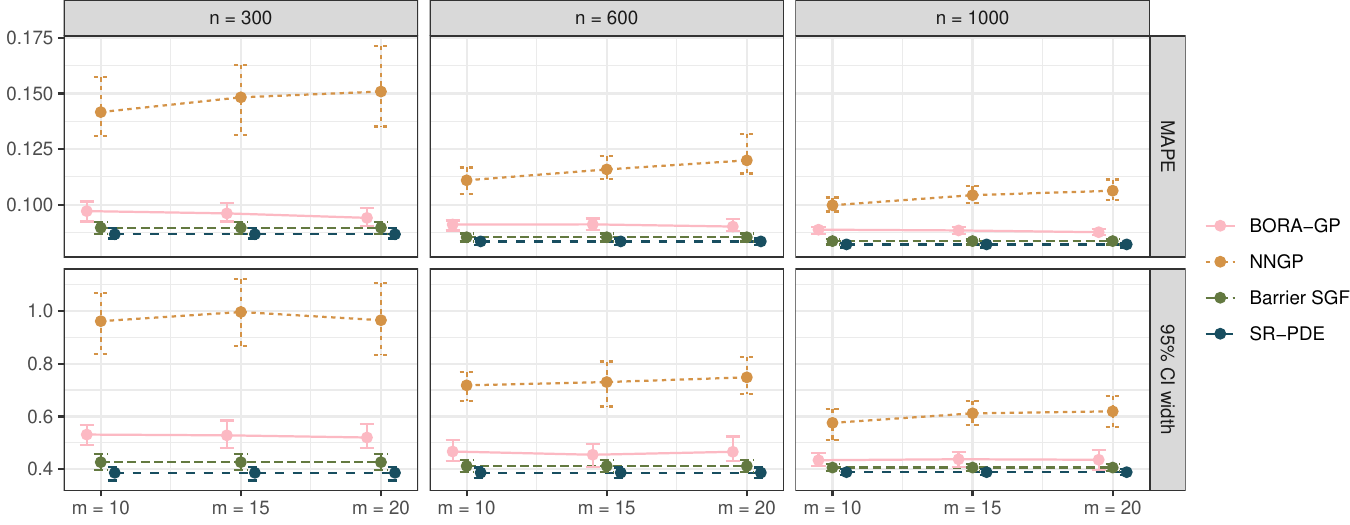}
\caption{Prediction performance of four methods by varying combinations of $n$ and $m$ in the horseshoe example. MAPE (top) and 95\% CI width (bottom) are illustrated as line plots where points indicate averages and line ranges are 0.025 and 0.975 quantiles.}
\label{suppfig:horseshoe_pred}
\end{figure}

\begin{landscape}
\begin{table}[ht]
    \centering
    \caption{Full simulation results of BORA-GP and NNGP from the modified horseshoe example. Average (standard error) of prediction performance metrics and posterior mean of model parameters of interest, and average computation time are provided over 30 synthetic data sets for each combination of $n$ and $m$.}
    \begin{tabular}{ccccccc}
    \hline
    &		\multicolumn{3}{c}{BORA-GP}	&	\multicolumn{3}{c}{NNGP} 	\\ \cmidrule(lr){2-4} \cmidrule(lr){5-7}
    	&	$m=10$	&	$m=15$	&	$m=20$ &	$m=10$	&	$m=15$	&	$m=20$ 	\\ \hline 
    $n=300$ & & & & & & \\
    RMSPE &0.122	(0.003)	&0.121	(0.003)	&0.118	(0.003)	&0.249	(0.028)	&0.259	(0.033)	&0.269	(0.039)	\\
    MAPE &0.097	(0.003)	& 0.096	(0.002)	& 0.094	(0.002)	&0.142	(0.007)	&0.148	(0.008)	&0.151	(0.010)		\\
    95\% CI coverage & 0.966	(0.009)	&0.966	(0.012)	&0.967	(0.010)	&0.966	(0.007)	&0.964	(0.010)	&\textbf{0.959}	(0.011)		\\
    95\% CI width & 0.531	(0.021)	&0.528	(0.030)	&0.520	(0.029)	&0.961	(0.069)	&0.996	(0.083)	&0.965	(0.078)		\\
    $\tau^2=0.01$ &0.009	(0.001)	&0.008	(0.001)	&0.009	(0.001)	&0.009	(0.001)	&0.009	(0.001)	&0.009	(0.001)		\\ 
    Preprocessing (sec.) & 4.122 & 6.807 & 7.813 & -- & -- & -- \\
    Runtime (sec.) & 76.903 & 186.324 & 301.499 & 84.872 & 184.974 & 287.132  \\
    Sec./iteration & 0.008 & 0.019 & 0.030 & 0.008 & 0.018 & 0.029  \\ \cmidrule(lr){1-7}
    $n=600$ & & & & & & \\
    RMSPE &0.114	(0.001)	&0.114	(0.002)	&0.113	(0.002)	&0.163	(0.015)	&0.171	(0.012)	&0.185	(0.022)	\\
    MAPE &0.091	(0.001)	&0.091	(0.001)	&0.090	(0.001)	&0.111	(0.004)	&0.116	(0.003)	&0.120	(0.005)	\\
    95\% CI coverage &0.955	(0.012)	&\textbf{0.949}	(0.013)	&0.957	(0.013)	&0.978	(0.006)	&0.972	(0.008)	&0.970	(0.008)	\\
    95\% CI width &0.467	(0.024)	&0.455	(0.026)	&0.466	(0.027)	&0.718	(0.034)	&0.730	(0.048)	&0.748	(0.042)	\\
    $\tau^2=0.01$ &0.009	(0.001)	&0.009	(0.001)	&0.009	(0.001)	&0.009	(0.001)	&0.009	(0.001)	&0.009	(0.001)	\\ 
    Preprocessing (sec.) & 4.069 & 5.493 & 6.915 & -- & -- & -- \\
    Runtime (sec.) & 87.756 & 174.800 & 309.809 & 77.865 & 180.948 & 292.920 \\
    Sec./iteration & 0.009 & 0.017 & 0.031 & 0.008 & 0.018 & 0.029  \\ \cmidrule(lr){1-7}
    $n=1000$ & & & & & & \\
    RMSPE &0.111	(0.001)	&0.111	(0.001)	&0.110	(0.001)	&0.131	(0.008)	&0.142	(0.008)	&0.146	(0.009)	\\
    MAPE &0.089	(0.001)	&0.088	(0.001)	&0.088	(0.001)	&0.100	(0.002)	&0.104	(0.002)	&0.106	(0.003)	\\
    95\% CI coverage &0.946	(0.009)	&\textbf{0.950}	(0.010)	&0.949	(0.012)	&0.973	(0.008)	&0.973	(0.006)	&0.970	(0.007)	\\
    95\% CI width &0.434	(0.014)	&0.438	(0.018)	&0.436	(0.021)	&0.575	(0.029)	&0.612	(0.024)	&0.619	(0.031)	\\
    $\tau^2=0.01$ &0.009	($<$0.001)	&0.009	($<$0.001)	&0.009	(0.001)	&0.008	($<$0.001)	&0.008	($<$0.001)	&0.008	(0.001)	\\ 
    Preprocessing (sec.) & 4.256 & 5.710 & 7.027 & -- & -- & -- \\
    Runtime (sec.) & 91.815 & 198.662 & 347.515 & 78.818 & 167.592 & 339.572 \\
    Sec./iteration & 0.009 & 0.020 & 0.035 & 0.008 & 0.017 & 0.034 \\
    \hline
    \end{tabular}
    \label{supptab:horseshoe_resall}
\end{table}
\end{landscape}

\begin{landscape}
\begin{table}[ht]
    \centering
    \caption{Full simulation results of Barrier SGF, Soap film smoother, SR-PDE, MDSdist, and ClosePD from the modified horseshoe example. Average (standard error) of prediction performance metrics and posterior mean of model parameters of interest, and average computation time are provided over 30 synthetic data sets for each $n$.}
    \begin{tabular}{cccccc}
    \hline
    & Barrier SGF & Soap film smoother & SR-PDE	& MDSdist & ClosePD \\ \hline
    $n=300$ & & & & & \\
    RMSPE & 0.113	(0.002)	& 0.131 (0.003) & \textbf{0.109} (0.002) & 0.255 (0.017) & 0.268 (0.019)\\
    MAPE & 0.090	(0.001)	& 0.101 (0.002) & \textbf{0.087} (0.001) & 0.116 (0.003) & 0.140 (0.006)\\
    95\% CI coverage & 0.941 (0.013)	& 0.590 (0.026) & 0.922 (0.014) & 0.999 (0.001) & 0.984 (0.007) \\
    95\% CI width & 0.426	(0.017)	& 0.199 (0.010) & 0.386 (0.016) & 1.389 (0.202) & 0.928 (0.030)\\
    $\tau^2=0.01$ & 0.009	(0.001)	& 0.015 (0.002) & 0.010 (0.001) & 0 (0) & 0 (0) \\ 
    Preprocessing (sec.) & 1.164 & -- & 0.018 & 111.475 & 165.374\\
    Runtime (sec.) & 4.546 & 13.223 & 6.587 & 6.256 & 5.474\\ \cmidrule(lr){1-6}
    $n=600$ & & & & & \\
    RMSPE & 0.107	(0.001)	& 0.125 (0.002) & \textbf{0.105} (0.001) & 0.253 (0.022) & 0.263 (0.016)\\
    MAPE & 0.085	(0.001)	& 0.096 (0.001) & \textbf{0.083} (0.001) & 0.115 (0.004) & 0.137 (0.006)\\
    95\% CI coverage & 0.943	(0.009)	& 0.552 (0.018) & 0.933 (0.010) & 0.998 (0.002) & 0.943 (0.010)\\
    95\% CI width & 0.412	(0.013)	& 0.173 (0.006) & 0.387 (0.012) & 1.130 (0.089) & 0.708 (0.028)\\
    $\tau^2=0.01$ & 0.009	(0.001) & 0.014 (0.001) & 0.010 (0.001) & 0 (0) & 0 (0)\\ 
    Preprocessing (sec.) & 1.461 & -- & 0.018 & 113.144 & 166.583\\
    Runtime (sec.) & 6.096 & 50.609 & 11.804 & 24.853 & 21.794\\ \cmidrule(lr){1-6}
    $n=1000$ & & & & &\\
    RMSPE & 0.105	(0.001)	& 0.122 (0.001) & \textbf{0.103} (0.001) & 0.261 (0.025) & 0.260 (0.018)\\
    MAPE & 0.084	(0.001)	& 0.094 (0.001) & \textbf{0.082} (0.001) & 0.117 (0.004) & 0.131 (0.005)\\
    95\% CI coverage & 0.945	(0.006)	& 0.525 (0.019) & 0.938 (0.006) & 0.993 (0.007) & 0.904 (0.021)\\
    95\% CI width & 0.406	(0.007)	& 0.158 (0.006) & 0.389 (0.006) & 0.949 (0.075) & 0.614 (0.047)\\
    $\tau^2=0.01$ & 0.009	($<$0.001) & 0.014 (0.001) & 0.010 ($<$0.001) & 0 (0) & 0 (0)\\ 
    Preprocessing (sec.) & 1.836 & -- & 0.018 & 115.749 & 169.209 \\
    Runtime (sec.) & 6.921 & 127.652 & 18.414 & 64.205 & 55.594 \\
    \hline
    \end{tabular}
    \label{supptab:horseshoe_resall_cont}
\end{table}
\end{landscape}

\subsection{Faults} \label{suppsec:faults}

Identical prior distributions are specified for both BORA-GP and NNGP: flat priors for $\beta_0$ and $\beta_1$ over $\R$, $\sigma^2 \sim IG(2,1)$, $\tau^2\sim IG(2,0.1)$, and $\nu$ is fixed at 1.5. A uniform prior is assigned for $\phi\sim Unif(2.240, 6.710)$ where the lower bound and the upper bound correspond to $\phi$ such that the correlation at 0.75 and 0.25 of the maximum distance in $\mc{D}$ is approximately 0.05. PC priors for Barrier SGF parameters are assigned so that the range parameter is larger than 0.9 and the spatial standard deviation is larger than 1.5 with prior probability 0.1. 

Since $\beta_0$ is not applicable in the implementation of SR-PDE through package \texttt{fdaPDE}, we use the mean-centered response in fitting SR-PDE. In addition, regression coefficients cannot be estimated in the current code for MDSdist and ClosePD. Therefore, we utilize ordinary least squares estimators for $\beta_0$ or $\beta_1$ in both methods. We further found that estimation of covariance parameters in MDSdist and ClosePD based on variograms was highly sensitive to initial values in this example. Hence, we used the true $\tau^2$ and $\sigma^2$ values as initial values with $\nu$ fixed at 1.5 in MDSdist and ClosePD. 

\begin{landscape}
\begin{table}[htbp]
    \centering
    \caption{Full simulation results of BORA-GP and NNGP when BORA-GP is the true data generating model in the domain with faults. Average (standard error) of prediction performance metrics and posterior mean of model parameters of interest, and average computation time are provided over 30 synthetic data sets.}
    \begin{tabular}{ccccccc}
    \hline
    &		\multicolumn{3}{c}{BORA-GP}	&	\multicolumn{3}{c}{NNGP}	\\ \cmidrule(lr){2-4} \cmidrule(lr){5-7}
    	&	$m=10$	&	$m=15$	&	$m=20$ &	$m=10$	&	$m=15$	&	$m=20$ 	\\ \hline 
    RMSPE &0.370 (0.007)	&0.369	(0.006)	&\textbf{0.368}	(0.006)	&0.393	(0.016)	&0.393	(0.016)	&0.393	(0.017)	\\
    MAPE &0.295	(0.006)	&\textbf{0.294}	(0.005)	&0.294	(0.005)	&0.310	(0.010)	&0.310	(0.010)	&0.309	(0.010)	\\
    95\% CI coverage &0.945	(0.011)	&0.947	(0.013)	&\textbf{0.948}	(0.011)	&0.956	(0.009)	&0.955	(0.007)	&0.954	(0.008)	\\
    95\% CI width &1.435	(0.061)	&1.435	(0.072)	&1.434	(0.070)	&1.593	(0.111)	&1.582	(0.111)	&1.572	(0.112)	\\
    $\beta_0=1$	&\textbf{1.032}	(0.420)	&1.094	(0.423)	&1.060	(0.422)	&1.059	(0.477)	&1.064	(0.445)	&1.033	(0.462)	\\
    $\beta_1=0.5$ &0.506	(0.023)	&0.506	(0.022)	&0.507	(0.023)	&0.506	(0.024)	&0.507	(0.024)	&0.507	(0.024)	\\
    $\sigma^2=1$ & 0.872	(0.237)	&0.846	(0.225)	&0.851	(0.243)	&0.806	(0.179)	&0.806	(0.176)	&0.852	(0.244)	\\
    $\tau^2=0.1$ &\textbf{0.100}	(0.010)	&0.100	(0.010)	&0.100	(0.010)	&0.119	(0.016)	&0.119	(0.016)	&0.121	(0.016)	\\
    $\phi=4$ &\textbf{4.458}	(0.613)	&4.555	(0.627)	&4.564	(0.668)	&5.271	(0.524)	&5.257	(0.543)	&5.061	(0.572)	\\
    Preprocessing (sec.) & 8.916 & 12.675 & 16.472 & -- & -- & -- \\
    Runtime (sec.) & 275.552 & 497.112 & 819.317 & 263.623 & 520.013 & 825.950 \\
    Sec./iteration & 0.018 & 0.033 & 0.055 & 0.018 & 0.035 & 0.055 \\
    \hline
    \end{tabular}
    \label{supptab:faults_resall}
\end{table}
\end{landscape}

\begin{landscape}
\begin{table}[htbp]
    \centering
    \caption{Full simulation results of Barrier SGF, Soap film smoother, SR-PDE, MDSdist, and ClosePD when BORA-GP is the true data generating model in the domain with faults. Average (standard error) of prediction performance metrics and posterior mean of model parameters of interest, and average computation time are provided over 30 synthetic data sets.}
    \begin{tabular}{cccccc}
    \hline
    &	Barrier SGF & Soap film smoother & SR-PDE & MDSdist & ClosePD \\ \hline 
    RMSPE &0.397	(0.018)	& 0.414 (0.020) & 0.412 (0.018) & 0.460 (0.107) & 0.525 (0.165)\\
    MAPE &0.312	(0.011)	& 0.324 (0.012) & 0.323 (0.012) & 0.365 (0.086) & 0.415 (0.127)\\
    95\% CI coverage &0.938	(0.010)	& 0.642 (0.028) & 0.948 (0.012) & 0.968 (0.035) & 0.921 (0.144)\\
    95\% CI width &1.476	(0.107)	& 0.737 (0.063) & 1.617 (0.141) & 2.267 (0.753) & 2.253 (0.957) \\
    $\beta_0=1$	&1.097 (0.352) & 1.100 (0.354) & -- & 1.100 (0.355) & 1.100 (0.355)\\
    $\beta_1=0.5$ &0.507	(0.023)	& 0.507 (0.026) & \textbf{0.504} (0.029) & 0.519 (0.048) & 0.519 (0.048)\\
    $\sigma^2=1$ &0.696	(0.226)	& -- & -- & \textbf{0.890} (0.501) & 0.799 (0.447) \\
    $\tau^2=0.1$ &0.118	(0.018) & 0.167 (0.032) & 0.171 (0.030)	& 0.255 (0.299) & 0.352 (0.323)\\
    $\phi=4$ &3.248	(0.562)	& -- & -- & $<$0.001 ($<$0.001) & $<$0.001 ($<$0.001)\\
    Preprocessing (sec.) & 7.082 & -- & 0.020 & 142.981 & 346.019\\
    Runtime (sec.) & 13.104 & 126.663 & 5.560 & 13.240 & 16.595\\
    \hline
    \end{tabular}
    \label{supptab:faults_resall_cont}
\end{table}
\end{landscape}

We run more simulations with increased widths of faults to further examine how Barrier SGF performs with thin barriers. Every other setting is identical to before. We found that Barrier SGF starts to show comparable performance to BORA-GP with 20 times thicker faults (see Table \ref{supptab:faults_thicker}). In this case, the barrier area occupies 0.15\% of the total area $[0,2]^2$. Even then, Barrier SGF still yields significantly nonzero residuals around the upper fault in some synthetic data sets as illustrated in Figure \ref{suppfig:faults_thicker_w}.
    
\begin{table}[htbp]
    \centering
    \caption{Average (standard error) of prediction performance metrics over 30 synthetic data sets in the domain with 20 times thicker faults.}
    \begin{tabular}{cccc}
    \hline
        & BORA-GP ($m=15$) & Barrier SGF \\ \hline
        RMSPE & \textbf{0.369} (0.006) & 0.375 (0.008) \\
        MAPE & \textbf{0.294} (0.005) & 0.298 (0.006) \\
        95\% CI coverage & \textbf{0.947} (0.013) & 0.942 (0.009) \\
        95\% CI width & 1.435 (0.072) & 1.427 (0.062) \\ \hline
    \end{tabular}
    \label{supptab:faults_thicker}
\end{table}

\begin{figure}
    \centering
    \includegraphics[scale = 0.72]{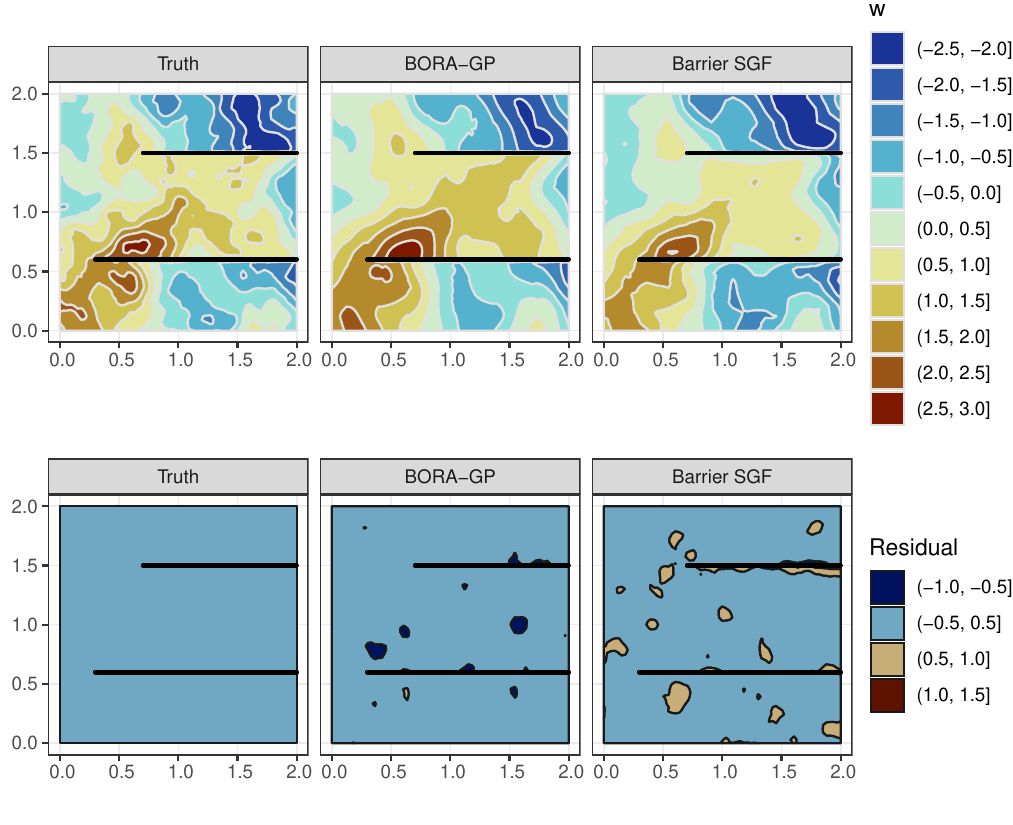}
    \caption{True and predicted $w(\bm{s})$ surfaces and residual ($w(\bm{s})-\hat{w}(\bm{s})$) surfaces by three methods from one (of 30) synthetic data set in the domain with 20 times thicker faults.}
    \label{suppfig:faults_thicker_w}
\end{figure}   

\section{Application to SSS in the Arctic Ocean} \label{suppsec:arctic}

We investigated convergence of MCMC on identifiable parameters using empirical diagnostic tools such as trace plots, running mean plots, and effective sample size (ESS). Running mean plots and trace plots are displayed in Figure \ref{suppfig:convergence}. The running mean plots show that we stopped iterations long after the running mean of the parameters of interest had stabilized. The trace plots show evidence of convergence to stationarity and adequate mixing. Running mean plots and trace plots of predicted $y(\bm{s})$ at 3,702 unobserved locations were also examined and suggested convergence to stationarity, but are omitted to save space. The \texttt{R} packages \texttt{posterior} and \texttt{mcmcse} are used to calculated univariate and multivariate ESS, respectively. According to \cite{vats_multivariate_2019}, 8,123 ESS is required to estimate three parameters, $\beta_0$, $\tau^2$, and $\sigma^2\phi^{2\nu}$, jointly with 95\% confidence and 5\% tolerance. Posterior samples of the three parameters estimate ESS to be 10,000 which exceeds the abovementioned ESS threshold. The minimum of univariate ESS estimates for predicted $y(\bm{s})$'s is as large as 10,717, and the maximum is 11,247.

\begin{figure}[!ht]
\centering
\includegraphics[width=0.85\textwidth]{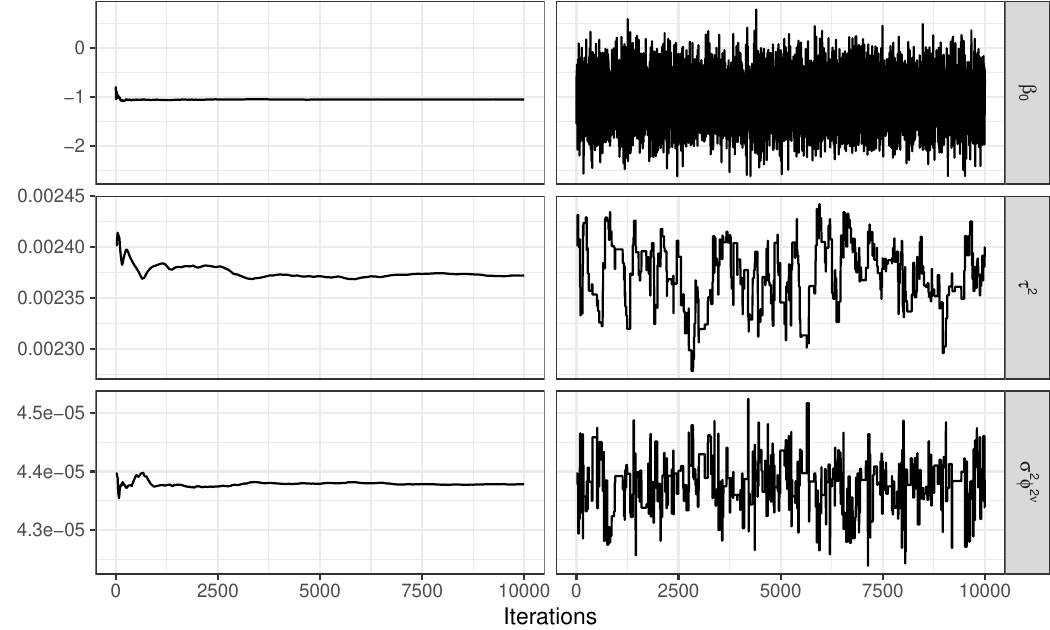}
\caption{Running mean plots on the left column and trace plots on the right column for $\beta_0$, $\tau^2$ and $\sigma^2\phi^{2\nu}$ in the application to SSS of the Arctic Ocean.}
\label{suppfig:convergence}
\end{figure}

\begin{figure}
\centering
\includegraphics[width=0.65\textwidth]{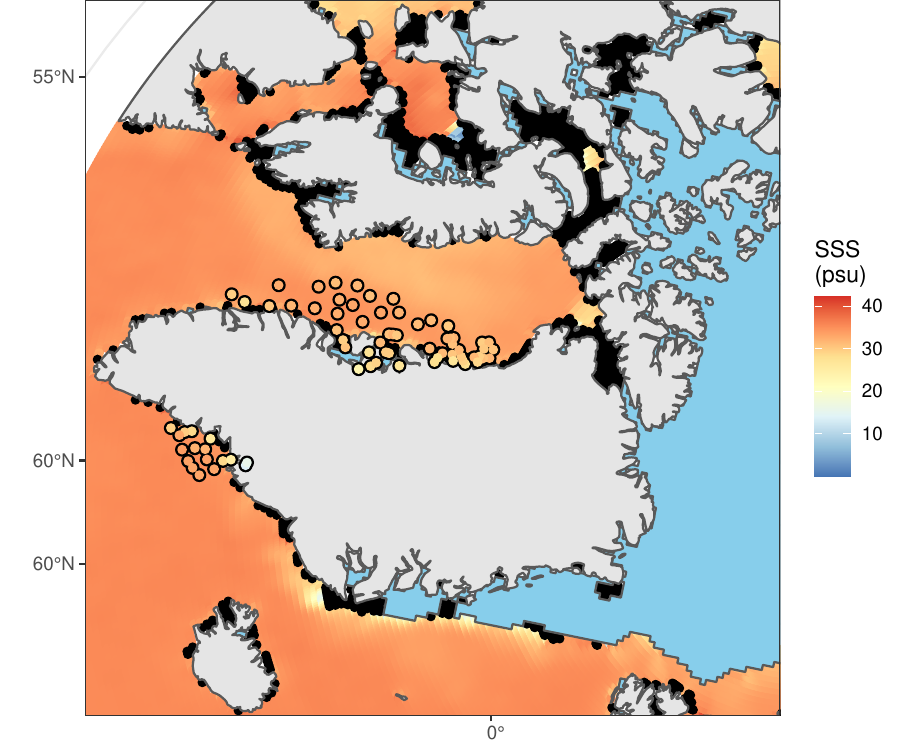}
\caption{OMG AXCTD measurements along the coast of Greenland as solid dots on top of SMAP SSS measurements. Locations to predict near the ice edge or land are drawn as black dots.}
\label{suppfig:omg}
\end{figure}

\section{Robustness of BORA-GP}
\label{suppsec:robust}

\subsection{Robustness to reference sets}
\label{suppsec:robust_to_R}

We examine how predictive surfaces change with different choices of the reference set. We reuse the modified horseshoe data in Figure \ref{fig:horseshoe} and fix $\tau^2 = 0$. Recall that $n=300$ locations are randomly selected and are not gridded. We run three BORA-GP models with $m=20$: (1) the reference set $\mc{R}$ is equal to the observation set $\mc{T}$, (2) the reference set $\mc{R}_1$ is a grid whose size $k=303$ is similar to $n$, and (3) the reference set $\mc{R}_2$ is a coarser grid with only 92 locations. As computational complexity depends on size of the reference set, researchers are rarely motivated to choose a reference set with $k$ much larger than $n$. 

Figure \ref{suppfig:horseshoe_freeS} exhibits almost indistinguishable predicted surfaces of $y(\bm{s})$ from the BORA-GP models with different reference sets. The predicted surfaces from BORA-GP with $\mc{R}=\mc{T}$ and $\mc{R}_1$ are highly similar. With the aid of contour lines, we observe that the prediction accuracy of BORA-GP with $\mc{R}_2$ somewhat deteriorates. In fact, RMSPEs in this specific data set are 0.022, 0.030, and 0.056 for BORA-GP with $\mc{R}=\mc{T}$, $\mc{R}_1$, and $\mc{R}_2$, respectively. The higher prediction errors from BORA-GP with $\mc{R}_2$ might be due to too coarse reference locations to accurately characterize correlations between other locations. Hence, we conclude that from a prediction perspective, the choice of the reference set may have little impact on inferences, unless $k$ is significantly less than $n$. This conclusion suggests to consider $\mc{T}$ as one of the most practical and sensible candidates of the reference set. 

\begin{figure}
    \centering
    \includegraphics[scale = 0.75]{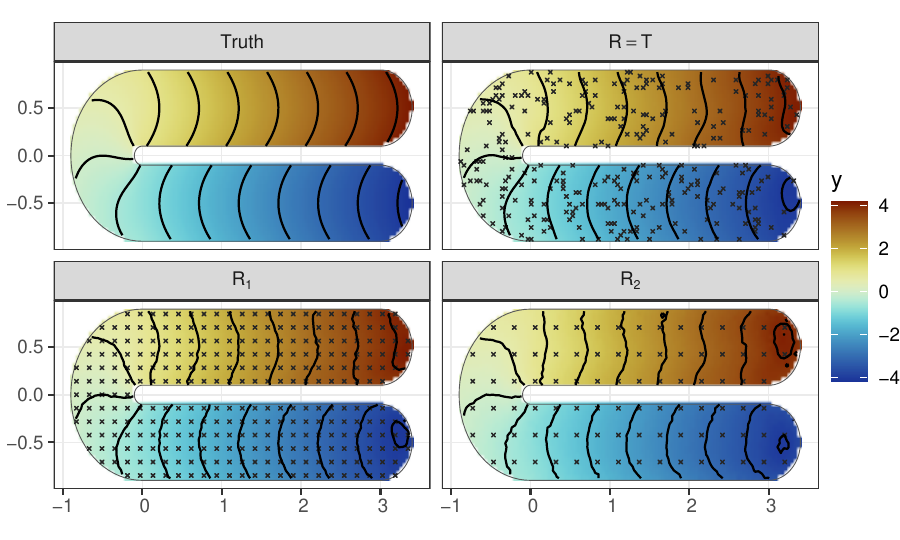}
    \caption{True and predicted $y(\bm{s})$ surfaces from BORA-GP models with various reference sets on the modified horseshoe data. Reference locations are marked with x.}
    \label{suppfig:horseshoe_freeS}
\end{figure}

As illustrated above, inferences on predicted values are robust to reference sets, which is also suggested in \cite{datta_hierarchical_2016}. Hence, we now focus on how correlations change by different reference sets in BORA-GP and NNGP in the SSS application in the Arctic Ocean. Two reference sets are exploited. One reference set is a finer grid, and the other is a coarser grid. Neither of them are equal to the observation set.  

First, Figure \ref{suppfig:arctic_cov_freeS} shows that the finer a gridded reference set is, the more neighbors overlap in BORA-GP and NNGP, letting the two methods have similar correlation patterns (top row). Second, comparing the bottom row of Figure \ref{fig:arctic_pred} with Figure \ref{suppfig:arctic_cov_freeS}, we observe that NNGP correlations may become more sensible if reference locations are uniform over the domain rather than set to irregularly observed locations. Nevertheless, NNGP still finds physically nonsensical neighbors across barriers and produces correlations inappropriately smoothed over the barriers. This issue is likely to persist in NNGP regardless of the design of the reference set if a domain is complex. Finally, BORA-GP correlations are robust to reference sets and continue to have higher values in the water outlined in an oval in Figure \ref{suppfig:arctic_cov_freeS} and keep correlations low in the Northwestern Passages across the barrier. 

\begin{figure}
    \centering
    \includegraphics[scale = 0.7]{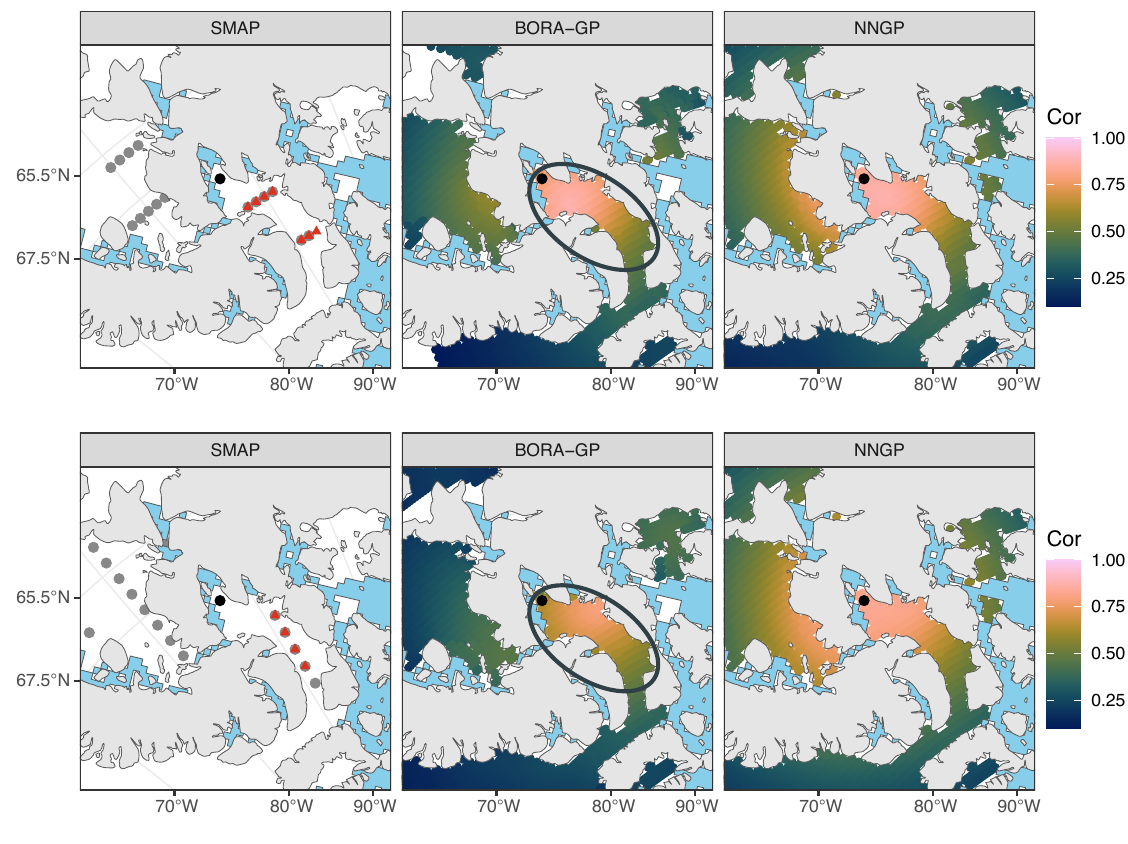}
    \caption{Each row displays correlations larger than 0.1 from a black point at (67.6$^{\circ}$N, 87$^{\circ}$W) by BORA-GP (middle) and NNGP (right) around the Canadian Arctic archipelago with a gridded reference set. On left panels, BORA-GP first-order neighbors and NNGP neighbors for the black point are marked in red triangles and gray circles, respectively.}
    \label{suppfig:arctic_cov_freeS}
\end{figure}
       
The robustness of BORA-GP covariance to reference sets is further examined in the domain with sliding doors in Section \ref{sec:boragp}. Figure \ref{suppfig:covsliding_diffRs} illustrates BORA-GP covariances with different choices of $\mc{R}$. The top two ((1,1) and (1,2)) panels are identical to Figure \ref{fig:covsliding}. Let the reference set $\mc{R}$ in panel (1,2) be denoted by $\mc{R}_{(1,2)}$. Other reference sets are chosen by omitting all reference locations at the opening along the $y$-axis from $\mc{R}_{(1,2)}$, coarsening $\mc{R}_{(1,2)}$ on the $y$-axis, and on both axes in (2,1), (2,2), and (3,1) panels, respectively. In panel (3,2), about the same number of locations as $\mc{R}_{(1,2)}$ are randomly selected as the reference set. Across BORA-GP covariances, the squeezed pattern by two sliding doors is evident regardless of the choice of $\mc{R}$. However, panel (2,1) illustrates that when there is a gap in $\mc{R}$ the peak contour ring widens to find physically sensible neighbors beyond the gap and that the highest covariances attained at three locations in the middle of the gap are lower than no-gap cases due to larger distances to BORA-GP neighbors. Panel (2,2) and (3,1) show that coarsening $\mc{R}$ in a certain axis makes contour lines slightly stretch along the axis. When reference locations are random in the domain as in (3,3) panel, which is often the case when the observed locations are chosen as reference locations, contour lines tend to deviate from circles or ovals. Nonetheless, nonstationary behaviors of covariance attributable to barriers persist, which supports choosing non-gridded observation locations as $\mc{R}$. 

\begin{figure}
    \centering
    \includegraphics[width = \textwidth]{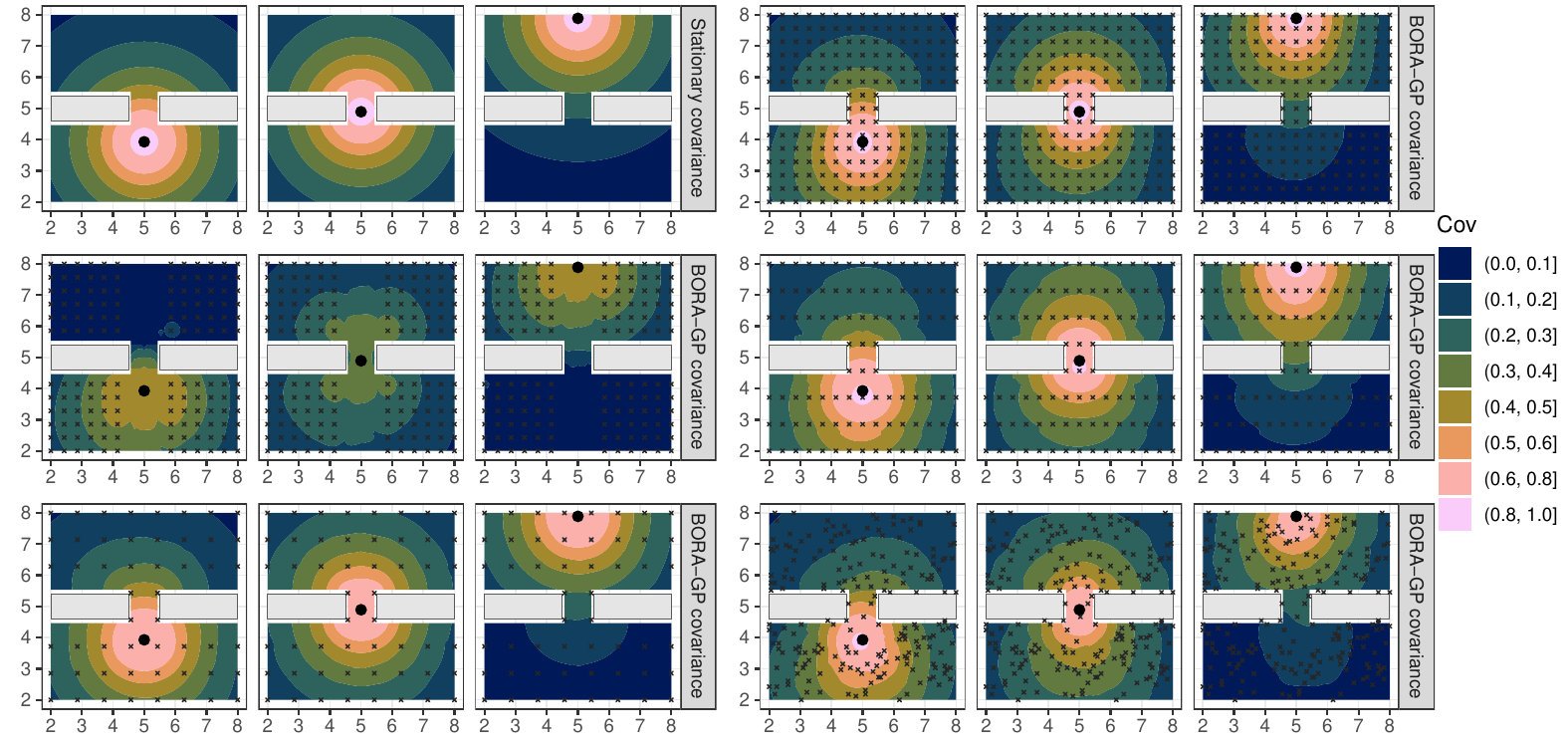}
    \caption{BORA-GP covariances are computed for three dots with 5 different sets of reference locations marked with x in a domain with a slight opening between sliding doors. The top-left panel presents stationary covariance for comparison.}
    \label{suppfig:covsliding_diffRs}
\end{figure}

\subsection{Robustness to ordering of reference locations}
\label{suppsec:robust_to_order}

With the Novaya Zemlya data in Section \ref{sec:novaya}, we compare three different orderings, in which locations are sorted in ascending order by $x$-, $y$-, and $x+y-$coordinates. Figure \ref{suppfig:novaya_flip} and Table \ref{supptab:novaya_flip} both suggest that ordering of $\mc{R}$ has negligible impact on prediction. 

\begin{figure}
    \centering
    \includegraphics[width = 0.8\textwidth]{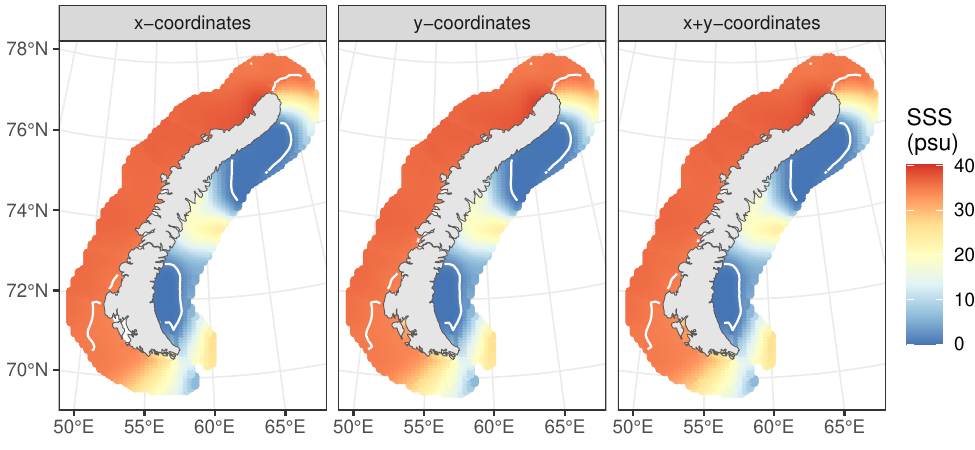}
    \caption{Predicted SSS bands around Novaya Zemlya on one randomly selected training-test split. White lines are contour lines at 1 and 35 psu.}
    \label{suppfig:novaya_flip}
\end{figure}

\begin{table}[htbp]
    \centering
    \caption{Prediction performance metrics are provided on one randomly selected training-test split of SSS data near Novaya Zemlya.}
    \begin{tabular}{cccc}
    \hline
         & $x$-coordinates & $y$-coordinates & $x+y$-coordinates\\ \hline 
   RMSPE & 0.181 & 0.182 & 0.182 \\
   MAPE  & 0.074 & 0.074 & 0.074 \\
   95\% CI coverage & 0.953 & 0.956 & 0.956 \\
   95\% CI width & 0.622 & 0.643 & 0.643 \\ \hline
   \end{tabular}
   \label{supptab:novaya_flip}
\end{table}
   
We also confirm the robustness of BORA-GP to orderings of $\mc{R}$ in the application to SSS in the Arctic Ocean. In Section \ref{sec:arctic}, we considered Transpolar Drift as one natural ordering in the Arctic Ocean and sorted observed locations aligned with directions of Transpolar Drift. In the current section, we reverse the ordering; the observed locations are sorted in the opposite direction of Transpolar Drift. 

We found that the ordering does not affect the posterior distribution of $\tau^2$ from BORA-GP or from NNGP; median and 95\% CI are consistently 0.0024 and (0.0023, 0.0024). Figure \ref{suppfig:arctic_flip_pred} also illustrates that predicted surfaces and covariance behaviors at a nonreference location do not present noticeable differences in reversing the ordering for BORA-GP or NNGP. In our examples, BORA-GP is robust to ordering of reference locations in terms of posterior summaries of parameters, prediction, and covariance.

\begin{figure}[htpb]
    \centering
    \includegraphics[scale = 0.7]{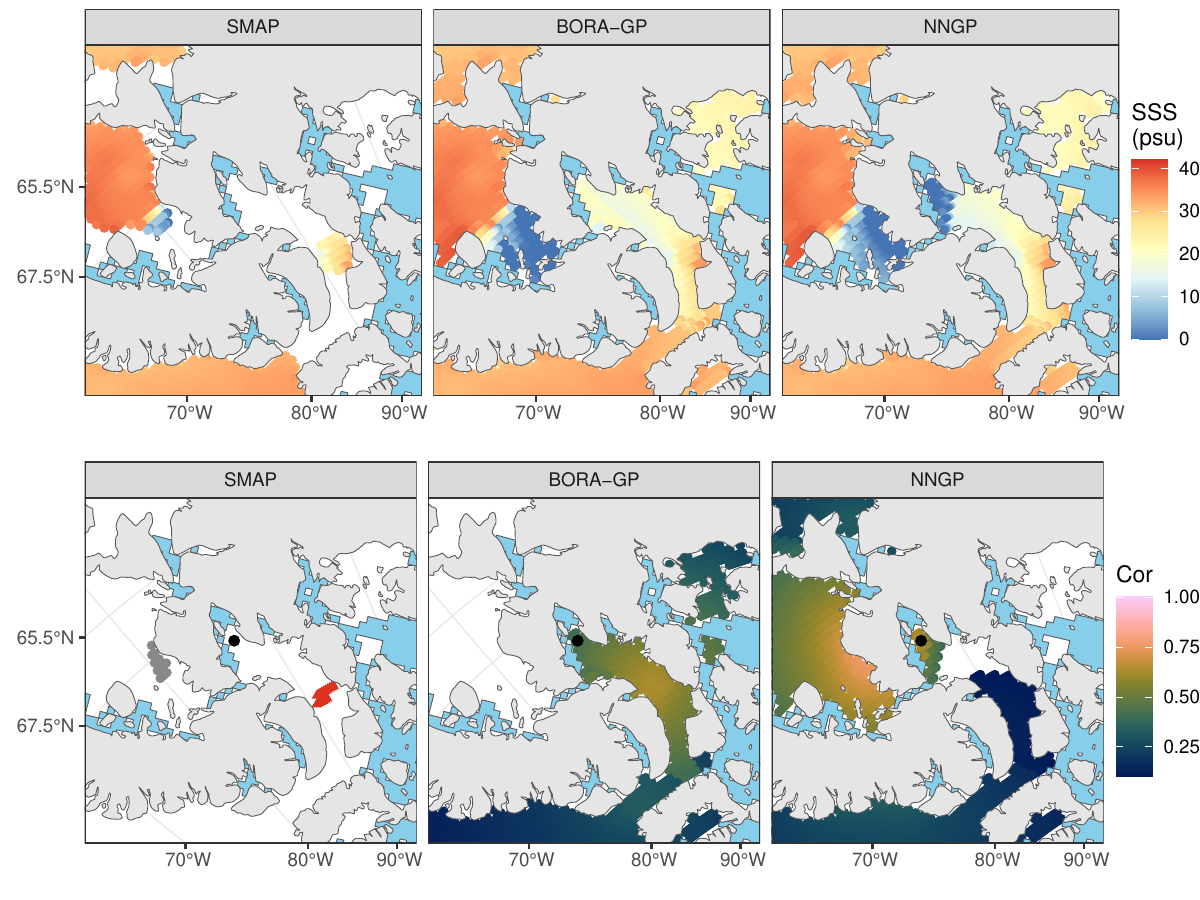}
    \caption{Top row shows the Canadian Arctic archipelago with SMAP SSS (left), predicted SSS by BORA-GP (center), by NNGP (right) in August, 2020. Same set of observed locations is used as Figure 12, but with the reversed ordering. The bottom row displays correlations $>0.1$ from a black point at (67.6$^{\circ}$N, 87$^{\circ}$W) by the two methods. On the bottom-left panel, BORA-GP neighbors and NNGP neighbors for the black point are marked in red triangles and gray circles, respectively.}
    \label{suppfig:arctic_flip_pred}
\end{figure}  
} \fi

\bibliographystyle{BJ_apalike}
\bibliography{bibliography}
	
\end{document}